\documentclass[prb,twocolumn]{revtex4}%

\usepackage{graphicx}
\usepackage{amsmath}
\usepackage{ulem}
\usepackage{color}

\newcommand{\be}{\begin{equation}}
\newcommand{\ee}{\end{equation}}
\newcommand{\bea}{\begin{eqnarray*}}
\newcommand{\eea}{\end{eqnarray*}}
\newcommand{\bean}{\begin{eqnarray}}
\newcommand{\eean}{\end{eqnarray}}

\begin{document}

\draft
\title{\bf Room-Temperature Pauli Spin Blockade and Current Rectification in 15-13-15 Armchair Graphene Nanoribbon Heterostructures}
\author{David M T Kuo}

\address{Department of Electrical Engineering and Department of Physics, National Central
University, Chungli, 32001 Taiwan}

\date{\today}

\begin{abstract}
In this study, we investigate the electronic structures of
13-11-13 and 15-13-15 armchair graphene nanoribbon (AGNR)
superlattices (SLs) using a tight-binding model. We demonstrate
that the conduction and valence subbands of 15-13-15 AGNR SLs can
be accurately described by the Su-Schrieffer-Heeger model, with
topologically protected interface states emerging at the junctions
between 15-AGNR and 13-AGNR segments. These interface states
enable the formation of quantum dot arrays with energy levels well
separated from bulk states, making them promising candidates for
high-temperature solid-state quantum processors. For 15-13-15
AGNRH segments, we observe both localized zigzag edge states and
topologically protected interface states under longitudinal
electric fields, with the latter providing efficient tunneling
channels in contrast to the less conductive edge states. We
further explore nonlinear charge transport through these interface
states under Pauli spin blockade, showing that tunneling current
spectra reveal charge stability diagrams and Coulomb blockade
oscillations, consistent with experimental findings in other
serial double quantum dot systems. Additionally, we examine the
impact of orbital offsets on tunneling current rectification and
demonstrate that significant current rectification is achieved
over a wide temperature range when level broadening is optimized.
These results highlight the potential of 15-13-15 AGNRHs for
robust spin-current conversion and applications in quantum
devices, offering advantages over other proposed structures due to
precise tunability of key parameters via bottom-up synthesis
techniques and the ease of two-gate electrode integration.
\end{abstract}

\maketitle

\section{Introduction}
Superposition and quantum entanglement are fundamental concepts in
quantum information science
[\onlinecite{DiVincenzo}-\onlinecite{ChuangIL}]. Quantum
processors are constructed from ideal quantum bits (qubits), which
are two-level, noise-resilient quantum systems
[\onlinecite{DiVincenzo}--\onlinecite{Bennett}]. A key requirement
for qubits is long quantum coherence times [\onlinecite{SHOR}]. As
a result, significant efforts have been devoted to identifying
physical systems that can host long-lived quantum states
[\onlinecite{Loss},\onlinecite{Nakamura}--\onlinecite{Kielpinski}].
Examples include quantum dots (QDs) [\onlinecite{Loss}],
superconducting junctions
[\onlinecite{Nakamura},\onlinecite{Makhlin}], double quantum dots
(DQDs) [\onlinecite{Ono},\onlinecite{vanderWiel}], trapped polar
molecules [\onlinecite{DeMille}], and ion traps
[\onlinecite{Kielpinski}]. Among them, charge quantum states in
QDs have demonstrated relatively long coherence times
[\onlinecite{Loss}]. While QDs offer promising scalability based
on modern semiconductor engineering, current realizations of
qubits are largely limited to DQD systems
[\onlinecite{Ono},\onlinecite{DiVincenzoDP}--\onlinecite{MaRL}].
However, reproducibility of identical DQD systems remains a major
challenge. Top-down lithographic techniques often suffer from
variability in QD size, inconsistent interdot separation, surface
defects, and contact interface issues. These imperfections lead to
fluctuations in critical parameters such as intradot and interdot
Coulomb interactions, interdot hopping strength, and electron
tunneling rates between QDs and electrodes. Such inconsistencies
hinder the scalability of QD-based qubits, which require highly
precise physical parameters for reliable qubit read/write
operations [\onlinecite{FrancescoR}--\onlinecite{MaRL}].

Since electron spin states typically exhibit longer coherence
times than charge states, they are considered more advantageous
for the physical realization of qubits. The discovery of graphene
with atomic thickness in 2004 [\onlinecite{Novoselovks}] has
motivated significant research efforts toward the development of
graphene-based spin qubits [\onlinecite{Trauzettel}]. Graphene
possesses exceptionally weak spin orbit coupling and negligible
hyperfine interaction, owing to the dominant presence of $^{12}$C
atoms with zero nuclear spin
[\onlinecite{AllenMT}--\onlinecite{Brotons}]. Recent advances in
bottom-up synthesis techniques have enabled the fabrication of
armchair graphene nanoribbon (AGNR) segments and AGNR
heterojunctions with atomic precision, effectively minimizing the
size fluctuations inherent in conventional quantum dots
[\onlinecite{Cai}--\onlinecite{SongST}]. Scanning tunneling
microscopy (STM) measurements on AGNR segments and 9-7-9 AGNR
heterojunctions have confirmed the existence of end zigzag edge
states and interface states, respectively
[\onlinecite{ChenYC}--\onlinecite{SongST}]. These topologically
protected localized states are highly promising candidates for
qubit implementations [\onlinecite{Albrecht}]. Furthermore,
AGNR-based superlattices (SLs) have shown great potential for
quantum processors [\onlinecite{BorsoiF}] and other quantum
devices [\onlinecite{WangHM}]. Consequently, it is of considerable
interest to investigate in-plane charge transport through GNRs
with edge sites coupled to electrodes, as this could provide
direct access to the observable properties of topological states
in graphene nanostructures
[\onlinecite{LlinasJP},\onlinecite{LiangGC}--\onlinecite{GuYan}].

To date, experimentally probing the electrical conductance and
tunneling current associated with charge transport through
topological states in AGNRs remains a significant challenge
[\onlinecite{LlinasJP},\onlinecite{ChenRS}]. One key reason is the
extremely short decay lengths of the end zigzag edge state wave
functions in 7-AGNR and 9-AGNR segments along the armchair edge
direction. Recently, we demonstrated that the zigzag edge states
in 13-AGNR segments exhibit much longer decay lengths compared to
those in 7-AGNR and 9-AGNR systems [\onlinecite{Kuo2}]. This
characteristic enhances the exchange interaction in graphene-based
serial double quantum dots (SDQDs)
[\onlinecite{Golor},\onlinecite{ChenCC}]. However, nonlinear
transport properties of 13-AGNR heterostructures (AGNRHs) have
been rarely explored [\onlinecite{LlinasJP},\onlinecite{DJRizzo}].
The main objective of this study is to investigate the electrical
conductance and tunneling current arising from topologically
protected states in AGNRHs incorporating 13-AGNR segments, as
illustrated in Figs. 1(a) and 1(b). These AGNRHs are constructed
from AGNR segments with differing widths $N$. When the zigzag edge
sites of a semiconductor/semiconductor/semiconductor AGNRH are
coupled to electrodes, as shown in Fig. 1(b), the system behaves
as an SDQD in which each quantum dot supports only a single energy
level [see Fig. 1(c)]. Because the energy levels $E_L$ and $E_R$
associated with the topologically protected QDs are well isolated
from the continuum states, they are robust against thermal noise,
defects, and vacancies. Therefore, we aim to systematically
analyze the room-temperature current rectification of such
topologically protected SDQDs under the Pauli spin blockade
configuration [\onlinecite{Ono},\onlinecite{BanY}]. Our results
offer valuable insight for the development of room-temperature
spin-current rectification and conversion technologies
[\onlinecite{BanY}].

\begin{figure}[h]
\centering
\includegraphics[trim=1.cm 0cm 1.cm 0cm,clip,angle=0,scale=0.3]{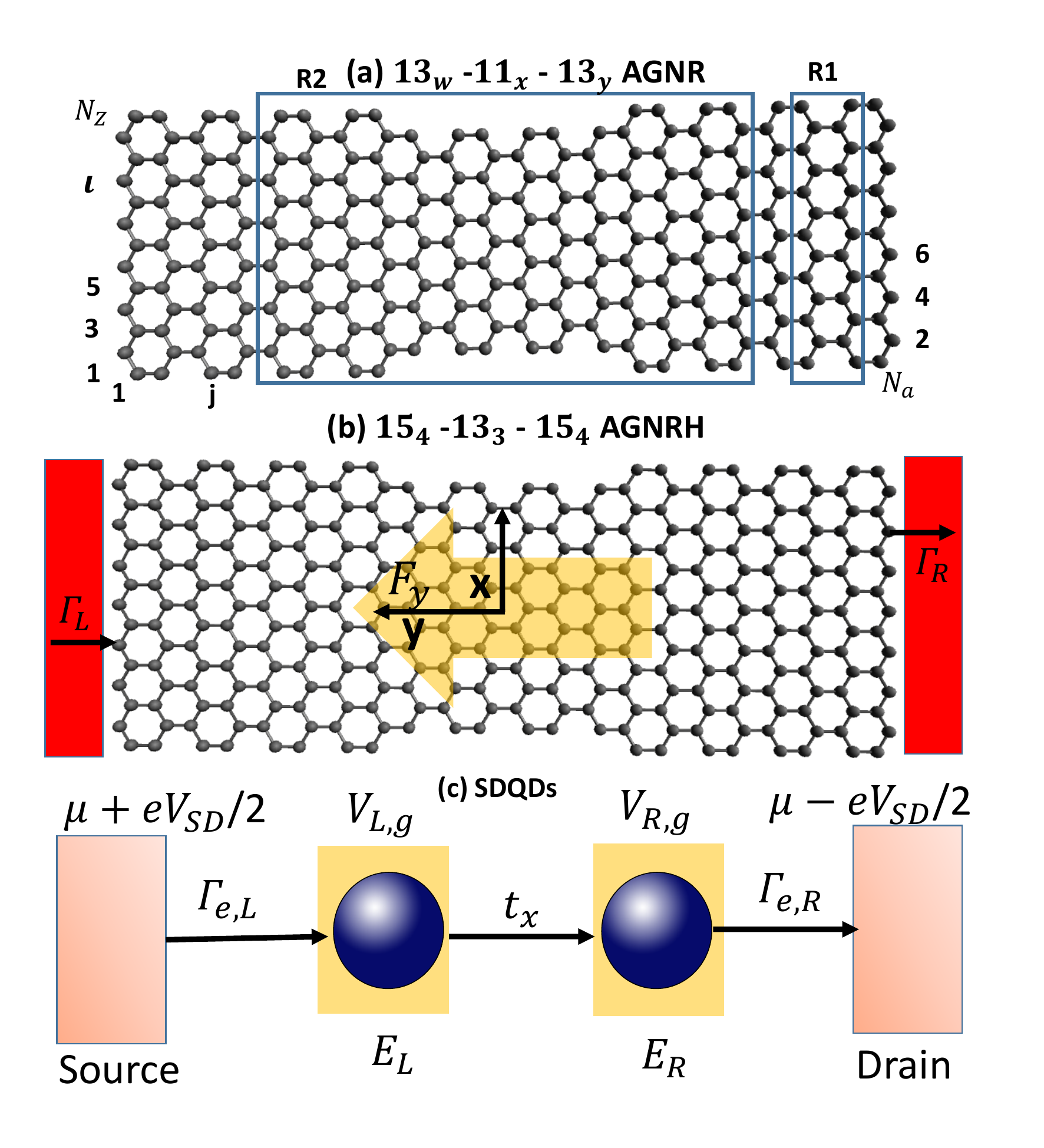}
\caption{ Schematic diagrams of armchair graphene nanoribbon
heterostructures (AGNRHs). (a) $13_{w}-11_{x}-13_{y}$ AGNRH, where
the indices $w$, $x$, and $y$ denote the lengths (in unit cells,
u.c.) of the respective AGNR segments. The unit cell of the AGNR
is labeled by $R_1$, and $R_2$ indicates the super unit cell
$13_2-11_3-13_2$ of the AGNR superlattice (AGSL). (b)
$15_4-13_3-15_4$ AGNRH with zigzag edge sites coupled to the
electrodes. Here, $\Gamma_{L}$ and $\Gamma_R$ represent the
electron tunneling rates between the left (right) electrode and
the atoms at the corresponding zigzag edge. (c) Schematic diagram
of the serial double quantum dot (SDQD) formed by the
topologically protected interface states of the AGNRHs. The SDQD,
characterized by effective tunneling rates $\Gamma_{e,L}$ and
$\Gamma_{e,R}$, is controlled by two gate voltages, $V_{L,g}$ and
$V_{R,g}$, as well as an applied voltage $V_{SD}$ between the
source and drain electrodes.}
\end{figure}

\section{Calculation method}
The system Hamiltonian, depicted in Fig.~1(b), is expressed as:
$H=H_0+H_{GNR}$, where $H_0$ represents the Hamiltonian of the
leads and $H_{GNR}$ describes the AGNRH. The Hamiltonian $H_0$ is
written as:

\begin{small}
\begin{eqnarray}
H_0& = &\sum_{k,\sigma} \epsilon_k
a^{\dagger}_{k,\sigma}a_{k,\sigma}+
\sum_{k,\sigma} \epsilon_k b^{\dagger}_{k,\sigma}b_{k,\sigma}\\
\nonumber &+&\sum_{\ell,k,\sigma}
V^L_{k,\ell,j}d^{\dagger}_{\ell,j,\sigma}a_{k,\sigma}
+\sum_{\ell,k,\sigma}V^R_{k,\ell,j}d^{\dagger}_{\ell,j,\sigma}b_{k,\sigma}
+ h.c.,
\end{eqnarray}
\end{small}
where the first two terms describe the free electrons in the left
and right electrodes. $a^{\dagger}_{k,\sigma}$
($b^{\dagger}_{k,\sigma}$) creates an electron of momentum $k$ and
spin $\sigma$ in the left (right) electrode, with energy
$\epsilon_k$. The terms $V^L_{k,\ell,j=1}$ ($V^R_{k,\ell,j=N_a}$)
describe the coupling between the left (right) lead and its
adjacent atom in the $\ell$-th row. The electronic states of the
GNR are described by a tight-binding model with one $p_z$ orbital
per atomic site [\onlinecite{Nakada}-\onlinecite{Wakabayashi2}].
The Hamiltonian $H_{GNR}$ for the AGNRH is:
\begin{small}
\begin{eqnarray}
H_{GNR} &= &\sum_{\ell,j} E_{\ell,j,\sigma} d^{\dagger}_{\ell,j,\sigma}d_{\ell,j,\sigma}\\
\nonumber&-& \sum_{\ell,j}\sum_{\ell',j'} t_{(\ell,j),(\ell', j')}
d^{\dagger}_{\ell,j,\sigma} d_{\ell',j',\sigma} + h.c,
\end{eqnarray}
\end{small}
where { $E_{\ell,j}$} is the on-site energy for the $p_z$ orbital
in the ${\ell}$-th row and $j$-th column. Here, the spin-orbit
interaction is neglected in this model.
$d^{\dagger}_{\ell,j,\sigma} (d_{\ell,j,\sigma})$ creates
(destroys) one electron at the atom site labeled by ($\ell$,$j$).
$t_{(\ell,j),(\ell', j')}$ describes the electron hopping energy
from site ($\ell$,$j$) to site ($\ell'$,$j'$). The tight-binding
parameters for the GNR are: $E_{\ell,j}=0$ for the on-site energy
and $t_{(\ell,j),(\ell',j')}=t_{pp\pi}=2.7$ eV for nearest
-neighbor hopping. Additionally, the effect of electric field
$F_y$ is included by the electric potential $U_y = e F_y y$ on
$E_{\ell,j}$, where $F_y = V_y/L_a$, with $V_y$ being the applied
voltage and $L_a$ the length of the AGNRH.

To study the transport properties of the AGNRH junction connected
to electrodes in the linear response region, we calculate the
electrical conductance as:

\begin{equation}
G_e=\frac{2e^2}{h}\int d\varepsilon~ {\cal
T}(\varepsilon)\frac{\partial f(\varepsilon)}{\partial \mu},
\end{equation}
where $f(\varepsilon)=1/(exp^{(\varepsilon-\mu)/k_BT}+1)$ is the
Fermi distribution function of electrodes at equilibrium
temperature $T$ and chemical potential $\mu$. $e$, $h$, and $k_B$
denote the electron charge, the Planck's constant, and the
Boltzmann constant, respectively. As seen in Eq. (3), the
transmission coefficient ${\cal T}(\varepsilon)$ plays a
significant role for electron transport during the ballistic
process.

The transmission coefficient ${\cal T}(\varepsilon)$  is
calculated using the following formula: ${\cal T}(\varepsilon) =
4Tr[\Gamma_{L}(\varepsilon)G^{r}(\varepsilon)\Gamma_{R}(\varepsilon)G^{a}(\varepsilon)]$
, where $\Gamma_{L}(\varepsilon)$ and $\Gamma_{R}(\varepsilon)$
are the tunneling rates (in energy units) at the left and right
leads, respectively, and $G^{r}(\varepsilon)$ and
$G^{a}(\varepsilon)$ are the retarded and advanced Green's
functions of the GNR, respectively
[\onlinecite{Kuo1},\onlinecite{SunQF}]. In the tight-binding
model, $\Gamma_{\alpha}(\varepsilon)$ and Green's functions are
matrices. The expression for $\Gamma_{L(R)}(\varepsilon)$ is
derived from the imaginary part of the self-energies, denoted as
$\Sigma^r_{L(R)}(\varepsilon)$, and is given by
$\Gamma_{L(R)}(\varepsilon)=-\text{Im}(\Sigma^r_{L(R)}(\varepsilon))=\pi\sum_k|V^{L(R)}_{k,\ell,j=1(N_a)}|^2\delta(\varepsilon-\epsilon_k)$.
For simplicity, we adopt the wide-band limitation for the
tunneling rates, $\Gamma_{L(R)}(\varepsilon)$, assuming that these
rates are energy-independent, and denote them as $\Gamma_{L(R)}$.
At zero temperature, the electrical conductance is given by
$G_e(\mu)=\frac{2e^2}{h}{\cal T}(\mu)$[\onlinecite{Mangnus}]. We
developed a Fortran computational code to calculate the electronic
structures described by Eq. (2) and the electrical conductance
given in Eq. (3).

\section{Results and discussion}
\subsection{Electronic structures of AGNR superlattices}
The electronic structures of AGNRs are determined by their $N_z$
-atom widths, where $N_z = 3p$, $3p+1$ and $3p+2$ with integer
number of $p$. When $N_z = 3p$ and $3p+1$, AGNRs are semiconductor
phases. When $N_z = 3p+2$, AGNRs are metallic phases
[\onlinecite{Nakada}-\onlinecite{Wakabayashi2}]. Therefore, the
structures of $13-11-13$ and $15-13-15$ AGNR heterostructures
illustrated in Fig. 1(a) and 1(b) can be regarded as
semiconductor/metal/semiconductor and
semiconductor/semiconductor/semiconductor junctions. To depict the
electronic structures of AGNR superlattice (AGSLs) with above two
configurations, we present the electronic structures of AGSLs
shown in Fig. 2. These AGSLs show the semiconductor phases even
for the semiconductor/metal/semiconductor SLs. As seen in Fig.
2(a) and 2(b), the bandwidths of conduction and valence subbands
are reduced when $x$ is tuned from $x = 5$ to $x = 10$ for
$13_2-11_x-13_2$ AGSLs. Addition, the band-gap ($E_g = E_c-E_v$)
between the conduction and valence subbands is reduced with
increasing $x$. For example, we have $E_g = 0.454$~eV and $E_g =
0.27$~eV for $x = 5$ and $x = 10$, respectively. Unlike
$13_2-11_x-13_2$ AGSLs, the band-gap of $15_2-13_x-15_2$ is
enhanced with increasing $x$ values. Such a behavior difference,
it is mainly attributed to the formation of conduction and valence
subbands resulting from topological interface states in the
$15-13-15$. The subband structures of $E_{c,SSH}$ and $E_{v,SSH}$
are well illustrated by the Su-Schrieffer-Heeger (SSH) model
[\onlinecite{SSH}--\onlinecite{ObanaD}], which provides an
analytical expression of $E_{c(v)}(k_y) = \pm
\sqrt{t^2_x+t^2_w-2t_xt_w~cos(k_y(\pi/L))}$. Here, $t_x$ and $t_w$
represent the electron hopping strengths in the 13-AGNR segment
and 15-AGNR segment, respectively, while $L$ denotes the length of
the super unit cell. For $x = 5$, we have $t_w = 177.55$~meV and
$t_x = 104.9$~meV. For $x = 10$, we have $t_w = 161$~meV and $t_x
= 26.5$~meV. The reduction of $t_x$ indicates that the the wave
function overlap between the interface states becomes weak with
increasing 13-AGNR segment length.

\begin{figure}[h]
\centering
\includegraphics[angle=0,scale=0.3]{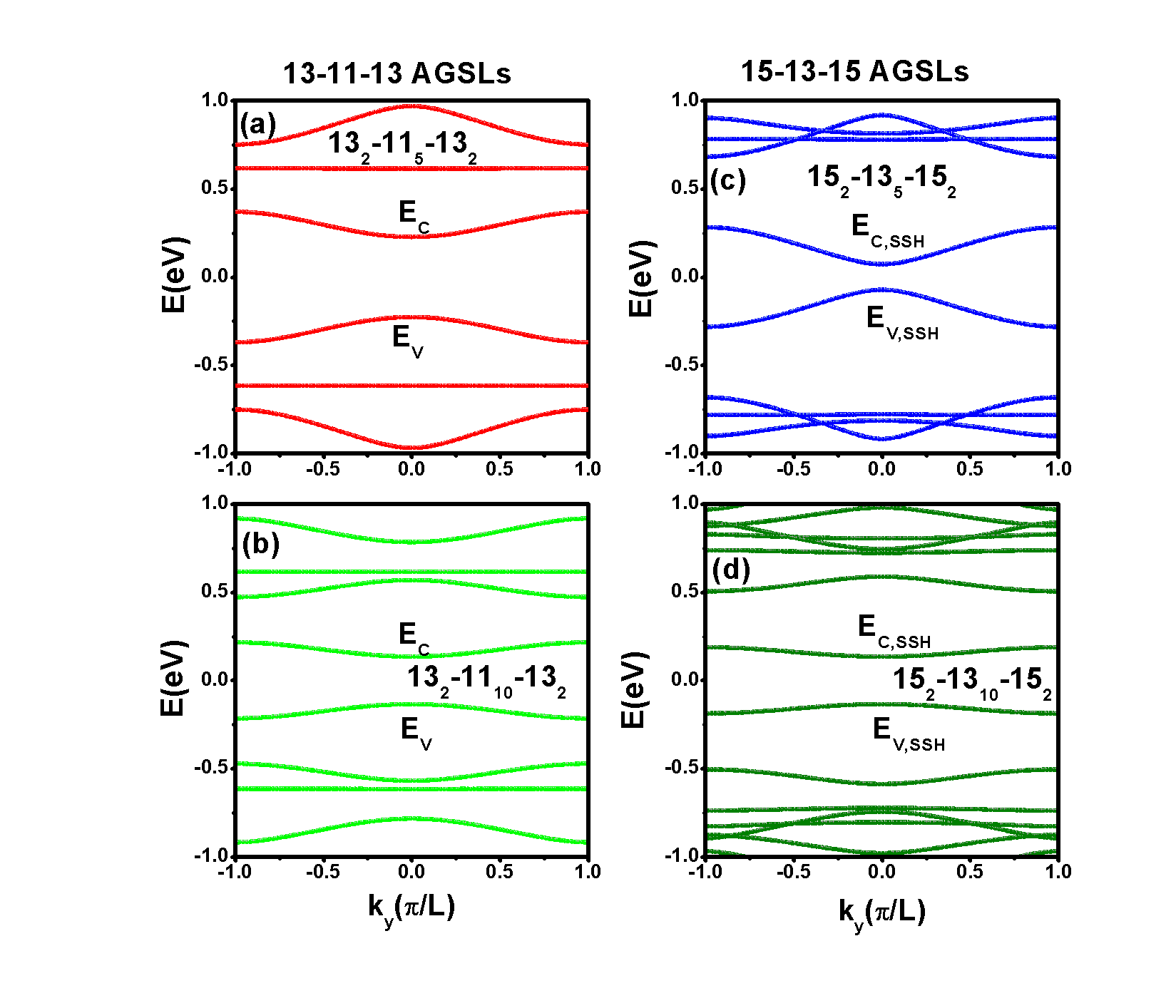}
\caption{Electronic subband structures of various AGNR
superlattice structures: (a) $13_2$-$11_5$-$13_2$, (b)
$13_2$-$11_{10}$-$13_2$, (c) $15_2$-$13_5$-$15_2$, and (d)
$15_2$-$13_{10}$-$15_2$. Here, $L$ denotes the length of each
superlattice. }
\end{figure}

\subsection{Energy levels of finite AGNRHs tuned by applied voltage}
Due to the challenges of realizing infinitely long AGNRs and AGSLs
using bottom-up synthesis techniques, the lengths of
experimentally fabricated AGNRHs are generally shorter than 20
nanometers [\onlinecite{ChenYC}--\onlinecite{SongST}]. In
addition, the 13-AGNR and 15-AGNR segments exhibit intriguing
topological multiple end zigzag edge states [\onlinecite{Zdetsis},
\onlinecite{Kuo2}]. Therefore, it is critical to investigate the
finite-size effects on the energy levels of AGNRHs under a uniform
electric field.

The calculated energy levels of $13_4$-$11_x$-$13_4$ AGNRHs are
presented in Figs. 3(a) and 3(b), which correspond to the cases of
$x = 5$ and $x = 10$ for $13_4$-$11_x$-$13_4$ AGNRHs,
respectively. Figs. 3(c) and 3(d) show the corresponding cases for
$15_4$-$13_x$-$15_4$ AGNRHs. The multiple end zigzag edge states
arise from AGNR segments with $N_z = 11, 13, 15, 17, 19$, and
$21$, among others. As a result, two left-end zigzag edge states
($|\Psi_{L,1}\rangle$ and $|\Psi_{L,2}\rangle$) and two right-end
zigzag edge states ($|\Psi_{R,1}\rangle$ and $|\Psi_{R,2}\rangle$)
are present at zero applied voltage. The left-end zigzag states
$|\Psi_{L,1(2)}\rangle$ and the right-end zigzag states
$|\Psi_{R,1(2)}\rangle$ form bonding and antibonding states,
denoted as $|\Psi_{B,1(2)}\rangle = |\Psi_{L,1(2)}\rangle +
|\Psi_{R,1(2)}\rangle$ and $|\Psi_{AB,1(2)}\rangle =
|\Psi_{L,1(2)}\rangle - |\Psi_{R,1(2)}\rangle$, respectively.
These four states correspond to the energy levels $\Sigma_{c1}$,
$\Sigma_{c2}$, $\Sigma_{v1}$, and $\Sigma_{v2}$, where the
subscripts "c" and "v" denote energy levels above and below the
charge neutrality point (CNP, $E=0$), respectively.

The application of an electric field induces a Stark effect that
lifts the near-degeneracy of these energy levels. Due to the
localized nature of their wave functions, the energy levels
exhibit a linear dependence on the applied voltage ($V_y$). As
shown in Figs.~3(a) and 3(b), the energy levels $\Sigma_{c1(2)}$
and $\Sigma_{v1(2)}$ are relatively insensitive to the variation
of the 11-AGNR segment length. In contrast, the energy levels
labeled $\Sigma_{c3}$ and $\Sigma_{v3}$, which originate from the
quantum confinement of the 11-AGNR segment, are more sensitive to
changes in segment length. These confined states exhibit a
bias-independent characteristic for $V_y \leq 0.5$~V.

In Figs.~3(c) and 3(d), the energy levels $\Sigma_{c3}$ and
$\Sigma_{v3}$ correspond to the bonding and antibonding states of
$|\Psi_{B(AB),I}\rangle = |\Psi_{L,I}\rangle \pm
|\Psi_{R,I}\rangle$, where $|\Psi_{L,I}\rangle$ and
$|\Psi_{R,I}\rangle$ are the left and right interface states of
the $15_w$-$13_x$-$15_y$ AGNRHs, respectively. In the absence of
an electric field, the energy levels of these topological
interface states are $\Sigma_{c3} = 88.22$~meV ($\Sigma_{v3} =
-88.22$~meV) and $\Sigma_{c3} = 20.46$~meV ($\Sigma_{v3} =
-20.46$~meV) for $x = 5$ and $x = 10$, respectively. As the length
of the 13-AGNR segment increases, the wave function overlap
between $|\Psi_{L,I}\rangle$ and $|\Psi_{R,I}\rangle$ weakens,
resulting in $\Sigma_{c3}$ and $\Sigma_{v3}$ moving closer to the
CNP. It is also noteworthy that the interface states can interact
with the bulk states, labeled by $E_{BC}$ and $E_{BV}$, leading to
the opening of a gap $\Delta$.

\begin{figure}[h]
\centering
\includegraphics[angle=0,scale=0.3]{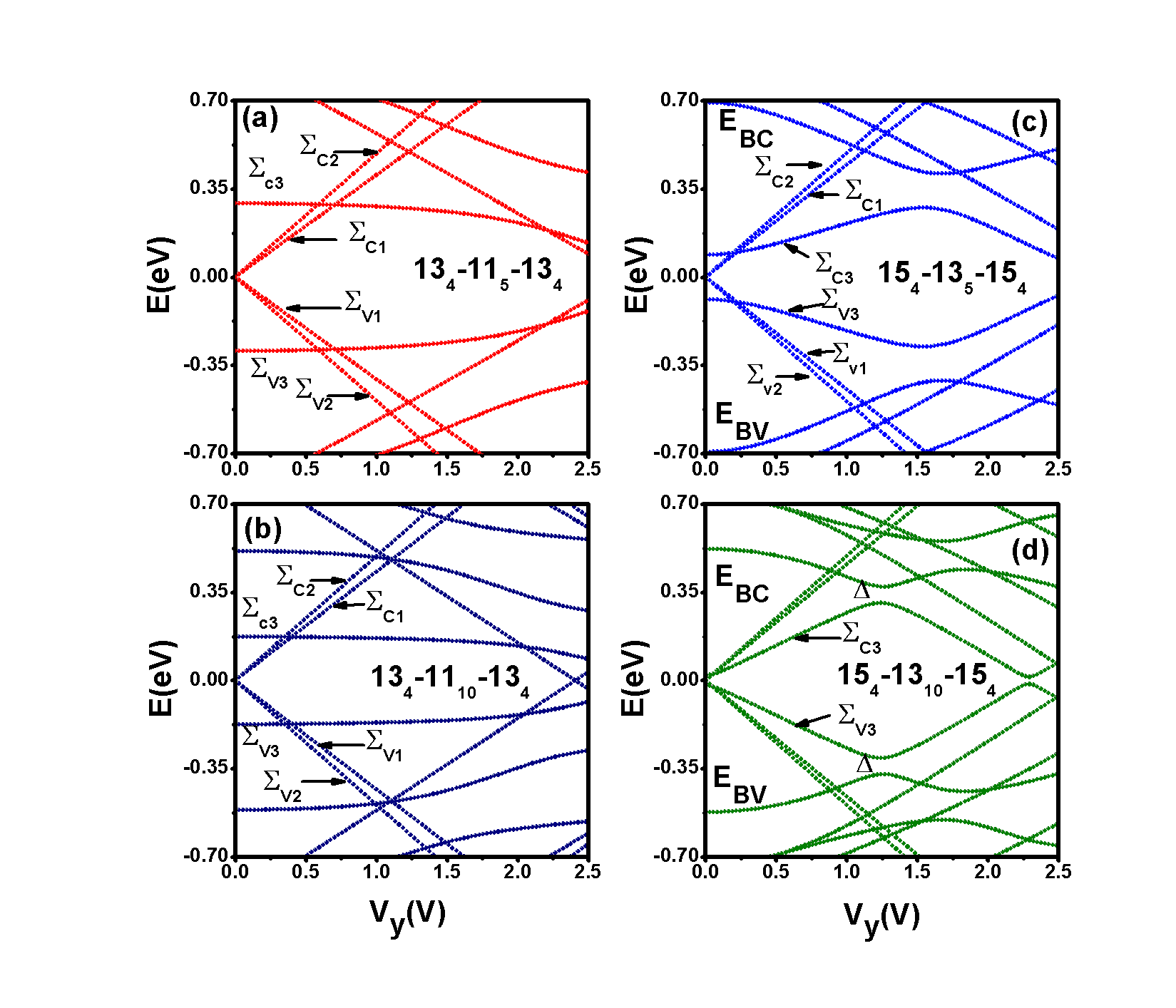}
\caption{Energy levels of AGNR heterostructures as functions of
the applied voltage $V_y$ for various structures: (a)
$13_4$-$11_5$-$13_4$, (b) $13_4$-$11_{10}$-$13_4$, (c)
$15_4$-$13_5$-$15_4$, and (d) $15_4$-$13_{10}$-$15_4$.}
\end{figure}

To further understand the multiple end zigzag edge states and
interface states of AGNRHs, we plot the charge densities
corresponding to $\Sigma_{c1}$, $\Sigma_{c2}$, and $\Sigma_{c3}$
for $13_4$-$11_5$-$13_4$ and $15_4$-$13_5$-$15_4$ AGNRHs at $V_y =
0$, as shown in Fig.~4. As seen in Fig.~4, the charge densities of
$\Sigma_{c1}$ and $\Sigma_{c2}$ are highly localized, primarily
concentrated at the sites of the end zigzag edge structures,
illustrating why these states are referred to as end zigzag edge
states. Comparing the charge density distributions between
$\Sigma_{c1}$ and $\Sigma_{c2}$, the wave function amplitude of
$\Sigma_{c2}$ decays more rapidly than that of $\Sigma_{c1}$. This
observation explains why the blue Stark shift of $\Sigma_{c2}$ is
larger than that of $\Sigma_{c1}$, as shown in Fig.~3.

For the $13$-$11$-$13$ AGNRH, the charge density of $\Sigma_{c3}$
is mainly confined within the 11-AGNR segment, although the
maximum charge density appears at the interface sites. Due to this
confinement, a significant Stark shift for $\Sigma_{c3}$ is
observed only when the applied voltage exceeds $V_y > 1$~V. In
contrast, for the $15_4$-$13_5$-$15_4$ AGNRH, the charge density
of $\Sigma_{c3}$ extends into the outer 15-AGNR segments and
decays rapidly along the armchair direction. This behavior also
explains why the Stark shift of $\Sigma_{c3}$ is smaller than that
of $\Sigma_{c1}$ and $\Sigma_{c2}$, as illustrated in Fig.~3(d).
Although both $15$-$13$-$15$ and $9$-$7$-$9$ AGNRHs belong to the
same class of topological heterostructures[\onlinecite{SanchoMP},
\onlinecite{Kuo4}], the wave functions of the interface states in
the $15$-$13$-$15$ AGNRH exhibit a much longer decay length
compared to those in the $9$-$7$-$9$ AGNRH segment (see Fig.
8(c)).

\begin{figure}[h]
\centering
\includegraphics[angle=0,scale=0.2]{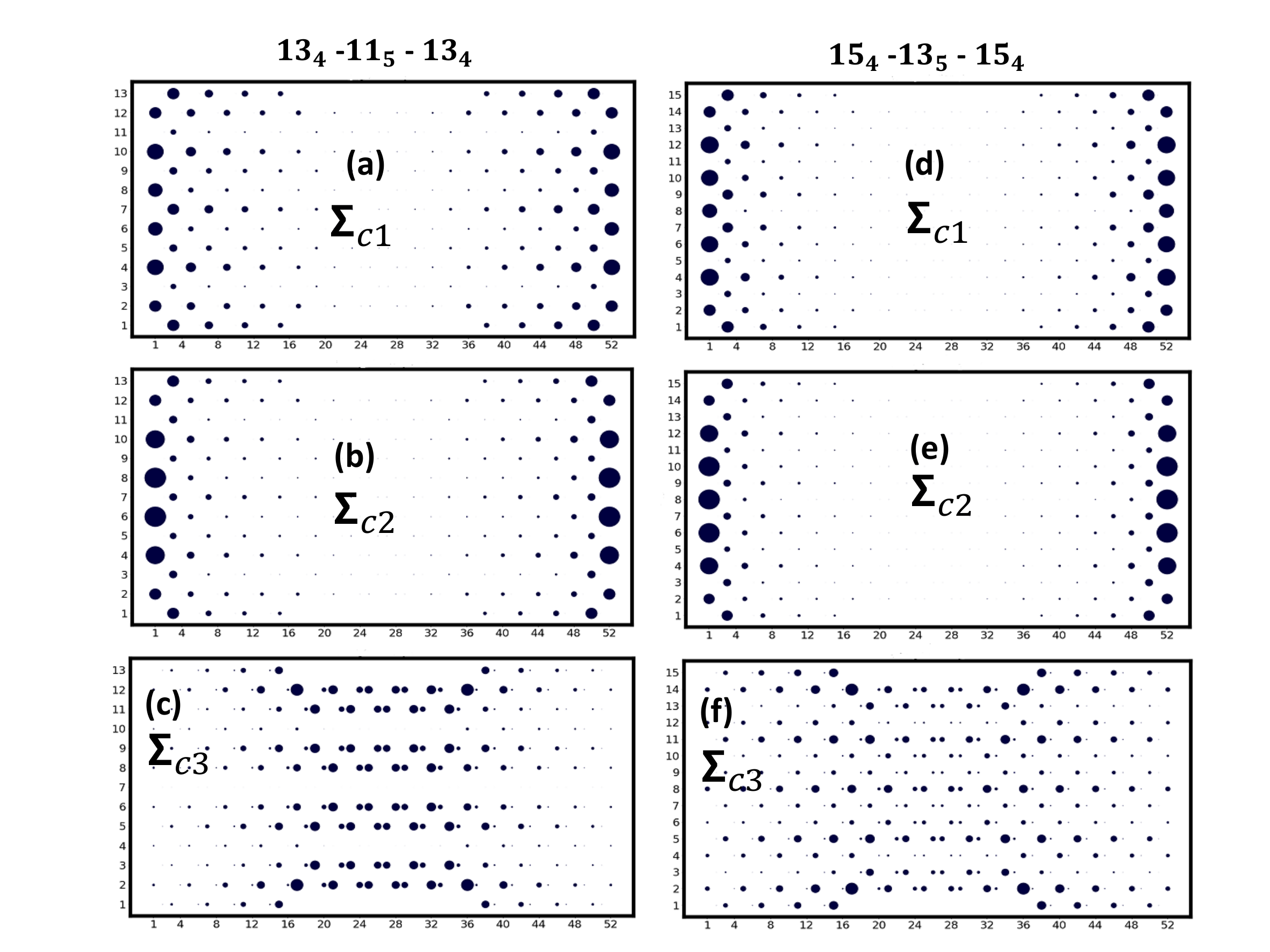}
\caption{Charge densities of topological states for
$13_4$-$11_5$-$13_4$ and $15_4$-$13_5$-$15_4$ AGNRHs. (a)
$\Sigma_{c1}$, (b) $\Sigma_{c2}$, and (c) $\Sigma_{c3}$ for the
$13_4$-$11_5$-$13_4$ AGNRH; (d) $\Sigma_{c1}$, (e) $\Sigma_{c2}$,
and (f) $\Sigma_{c3}$ for the $15_4$-$13_5$-$15_4$ AGNRH. The
radius of each circle represents the magnitude of the charge
density for the topological states.}
\end{figure}

\subsection{Electrical conductance of AGNRHs}
Although the topologically protected interface states of 9-7-9 and
7-9-7 AGNRHs have been experimentally confirmed via STM spectra,
such measurements do not reveal the in-plane charge transport
through these interface states
[\onlinecite{DRizzo}--\onlinecite{DJRizzo}]. To employ these
topological states in the implementation of quantum bits, it is
crucial to clarify their in-plane transport properties under
different contact geometries and materials. Based on Eq.~(3), we
present the electrical conductance as a function of chemical
potential $\mu$ for a $13_4$-$11_5$-$13_4$ AGNRH with zigzag edge
sites coupled to the electrodes, where $\Gamma_L = \Gamma_R =
\Gamma_t = 2.7$~eV in Fig. 5(a). The electrical conductance
spectra reveal features associated with the $\Sigma_{c3}$ and
$\Sigma_{v3}$ states, but do not showcase the localized states
$\Sigma_{c1}$, $\Sigma_{c2}$, $\Sigma_{v1}$, and $\Sigma_{v2}$.
For $\Gamma_t = t_{pp,\pi} = 2.7$~eV, the electrodes are regarded
as graphene line contacts.

To further reveal these end zigzag edge states, we calculate the
electrical conductance $G_e$ of $13_4$-$11_2$-$13_4$ AGNRHs with
shorter segments as a function of $\mu$ for various $\Gamma_t$
values. Specifically, we consider $\Gamma_t = 72$~meV, $54$~meV,
$36$~meV, $18$~meV, and $9$~meV, corresponding to Figs.~5(b),
5(c), 5(d), 5(e), and 5(f), respectively. As seen in Fig.~5(b), a
significant conductance peak $G_e(\mu = 0) = 0.3~G_0$ appears,
labeled as $\Sigma_0$, where $G_0 = \frac{2e^2}{h}$ denotes the
quantum of conductance. As $\Gamma_t$ decreases, the peak
associated with $\Sigma_0$ is significantly enhanced. For example,
$G_e$ approaches nearly one quantum conductance when $\Gamma_t =
18$~meV, while $\Sigma_0$ splits into two distinct peaks.

When $\Gamma_t = 9$~meV, we observe clear conductance features
corresponding to $\Sigma_{v1}$ and $\Sigma_{c1}$, but not to
$\Sigma_{v2}$ and $\Sigma_{c2}$. We confirm that these two peaks
arise from $\Sigma_{v1}$ and $\Sigma_{c1}$ by analyzing their
charge density distributions. It is worth noting that the results
in Fig.~5 do not include the real part of the self-energies
arising from the coupling between the electrodes and adjacent
carbon atoms. However, as the electrodes possess wide bands, the
real part of the self-energy is expected to be vanishingly small.
Nevertheless, interface issues between metallic electrodes and
GNRs may introduce additional contact problems, particularly
affecting the end zigzag edge states of AGNR segments
[\onlinecite{MatsudaY}].

\begin{figure}[h]
\centering
\includegraphics[angle=0,scale=0.3]{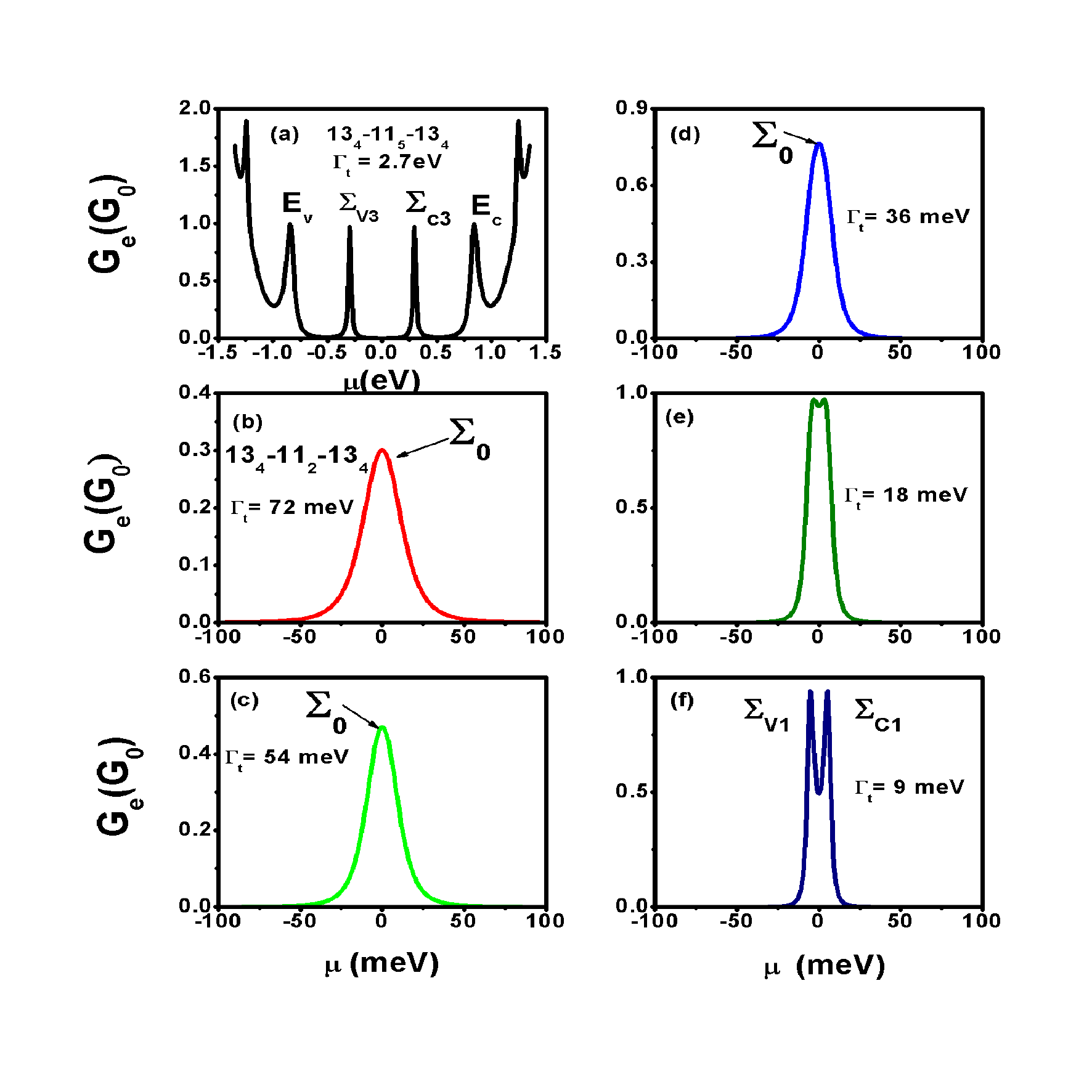}
\caption{(a) Electrical conductance $G_e$ of the
$13_4$-$11_5$-$13_4$ AGNRH with $\Gamma_t = 2.7$ eV and $N_a = 52$
($L_a = 5.4$~nm) as a function of chemical potential $\mu$ at zero
temperature. (b)-(f) Electrical conductance $G_e$ of the
$13_4$-$11_5$-$13_4$ AGNRH as functions of $\mu$ at zero
temperature for various values of $\Gamma_t$: (b) $\Gamma_t =
72$~meV, (c) $\Gamma_t = 54$~meV, (d) $\Gamma_t = 36$~meV, (e)
$\Gamma_t = 18$~meV, and (f) $\Gamma_t = 9$~meV. Symmetrical
tunneling rates are considered, with $\Gamma_L = \Gamma_R =
\Gamma_t$.}
\end{figure}

Next, we present the electrical conductance of
$15_4$-$13_x$-$15_4$ AGNRHs as a function of $\mu$ for various $x$
values at zero temperature and with $\Gamma_t = 2.7$~eV in Fig.~6,
where $x = 5, 6, 7, 8, 9$, and $10$ correspond to
Figs.~6(a)--6(f), respectively. The electrical conductance spectra
reveal clear features associated with the $\Sigma_{c3}$ and
$\Sigma_{v3}$ states for $x = 5$, as shown in Fig.~6(a). However,
when $x = 6$, the peaks corresponding to $\Sigma_{c3}$ and
$\Sigma_{v3}$ become unresolved in Fig.~6(b), while the maximum
electrical conductance reaches one quantum conductance ($G_e(\mu =
0) = G_0$). When $x = 7$, $G_e(\mu = 0)$ is reduced to below one
quantum conductance. As illustrated in Fig.~6, the conductance
peaks associated with $\Sigma_{c3}$ and $\Sigma_{v3}$ are
progressively suppressed as the $x$ value increases.

To further interpret these $G_e$ spectra, we employ the analytical
expression for the electrical conductance of a SDQD system without
Coulomb interactions, which describes charge transport through the
interface states of $15$-$13$-$15$ AGNRHs:

\begin{eqnarray}
& &G_e/G_0\nonumber \\
&=&\frac{4\Gamma_{e,L}t^2_{x}\Gamma_{e,R}}{|(\mu-E_L+i\Gamma_{e,L})(\mu-E_R+i\Gamma_{e,R})-t^2_{x}|^2},
\end{eqnarray}
where $\Gamma_{e,L} = \Gamma_{e,R} = \Gamma_{e,t}$ represents the
effective tunneling rate, $E_L = E_R = 0$ are the energy levels of
the left and right interface states, and $t_x$ denotes the
electron hopping strength between the two interface states.

The $t_x$ values are $88.22$~meV, $65.32$~meV, $48.64$~meV,
$36.36$~meV, $27.25$~meV, and $20.46$~meV for $x = 5, 6, 7, 8, 9$,
and $10$, respectively. According to Eq.~(4), the maximum
conductance $G_e(\mu = 0) = G_0$ occurs when $\Gamma_{e,t} = t_x$.
Therefore, in Fig.~6(b), we infer that the effective tunneling
rate of the interface states is $\Gamma_{e,t} = 65.32$~meV. The
value of $\Gamma_{e,t}$ is determined by the coupling strength
$\Gamma_t$ and the charge density distribution within the 15-AGNR
segments. Furthermore, Eq.~(4) also explains the observed
reduction in $G_e(\mu = 0)$, labeled as $\Sigma_0$, in
Figs.~6(c)--6(f).

\begin{figure}[h]
\centering
\includegraphics[angle=0,scale=0.3]{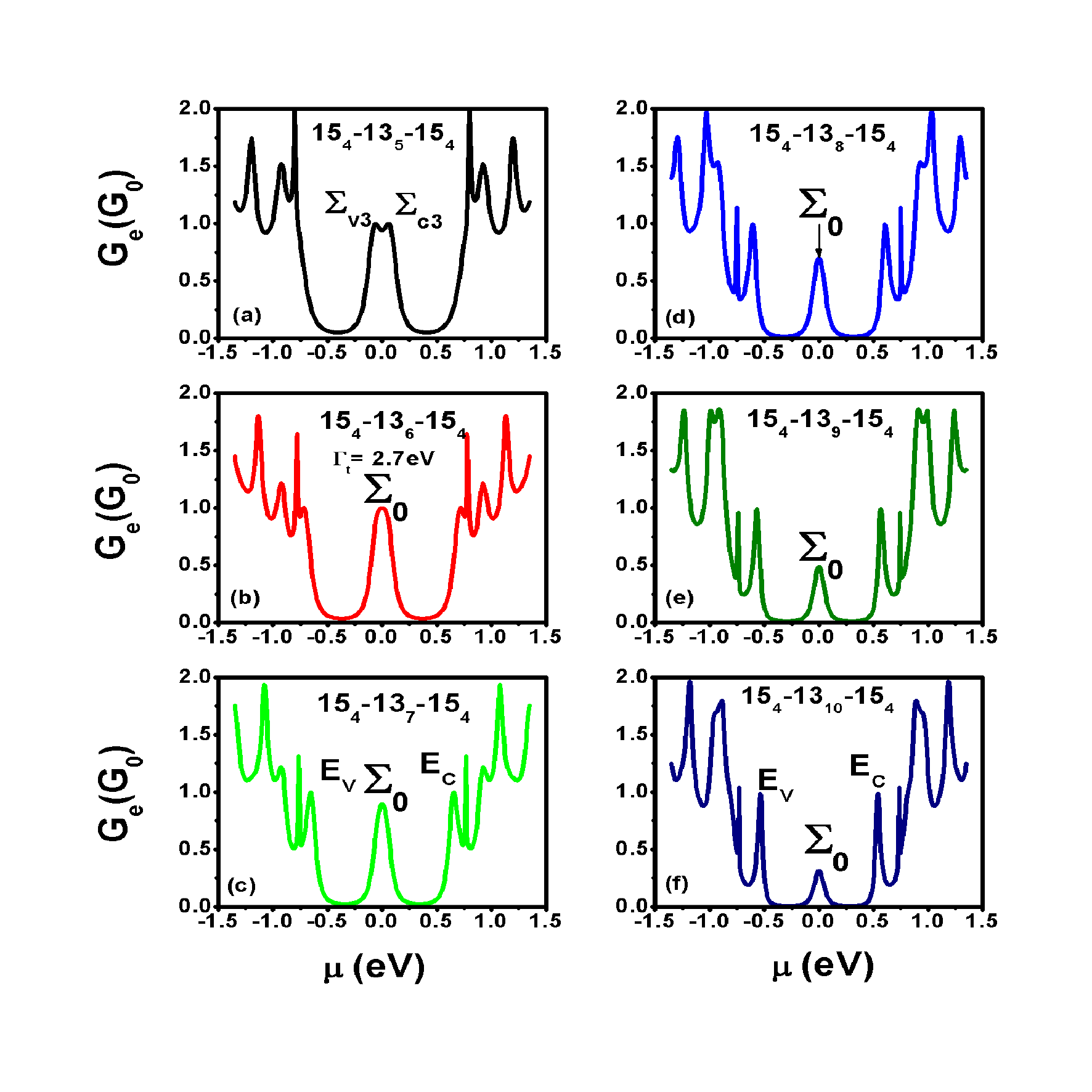}
\caption{Electrical conductance $G_e$ of the $15_4$-$13_x$-$15_4$
AGNRH with zigzag edge sites coupled to the electrodes for various
lengths of the 13-atom AGNR segments: (a) $x = 5$, (b) $x = 6$,
(c) $x = 7$, (d) $x = 8$, (e) $x = 9$, and (f) $x = 10$. The
tunneling rate is set to $\Gamma_t = 2.7$~eV.}
\end{figure}

In Figs.~5 and 6, we considered systems where the zigzag edge
sites are coupled to the electrodes. To further explore charge
transport through multiple end zigzag edge states, we present in
Fig.~7 the calculated electrical conductance $G_e$ of AGNRs where
the armchair edge sites are coupled to the electrodes, with
$\Gamma_t = 9$~meV. We investigate three AGNR structures: a
13-AGNR segment, a 15-AGNR segment, and a $15$-$13$-$15$ AGNRH
segment. As shown in Fig.~7(a), the $G_e$ spectra of the 13-AGNR
segment with $N_a = 52$ ($L_a = 5.4$~nm) exhibit peaks
corresponding to $\Sigma_{c1}$, $\Sigma_{v1}$, $\Sigma_{c2}$, and
$\Sigma_{v2}$, arising from multiple end zigzag edge states.
Notably, the conductance peak at $\mu = 0$ exceeds one quantum
conductance, which is attributed to the very small energy level
separation between $\Sigma_{c2}$ and $\Sigma_{v2}$. In contrast,
the energy separation between $\Sigma_{c1}$ and $\Sigma_{v1}$ is
relatively large, around $33.12$~meV.

In Fig.~7(b), we show the $G_e$ spectra of the 15-AGNR segment
with $N_a = 44$ ($L_a = 4.5$~nm). Similar to the 13-AGNR case,
$G_e(\mu = 0)$ exceeds one quantum conductance due to
contributions from the closely spaced energy levels of
$\Sigma_{c2}$ and $\Sigma_{v2}$. However, the peaks labeled by
$\Sigma_{c1}$ and $\Sigma_{v1}$ exhibit a very small energy
separation of $5.76$~meV, indicating that the wave functions of
the corresponding end zigzag edge states decay more rapidly in
15-AGNRs compared to 13-AGNRs.

In Fig.~7(c), only three peaks are observed in the $G_e$ spectra
of the $15_4$-$13_5$-$15_4$ AGNRH segment with $N_a = 52$. Two of
these peaks, labeled $\Sigma_{c3}$ and $\Sigma_{v3}$, originate
from the interface states of the $15$-$13$-$15$ heterostructure.
The central peak, labeled $\Sigma_0$, is associated with the
multiple end zigzag edge states. When $N_a = 52$, the four energy
levels $\Sigma_{c1}$, $\Sigma_{v1}$, $\Sigma_{c2}$, and
$\Sigma_{v2}$ become closely spaced and merge together. Overall,
when the armchair edge sites are coupled to the electrodes, the
transport behavior of $15$-$13$-$15$ AGNRHs resembles that of a
multiple quantum dot system~[\onlinecite{OtsukaT}].

\begin{figure}[h]
\centering
\includegraphics[angle=0,scale=0.3]{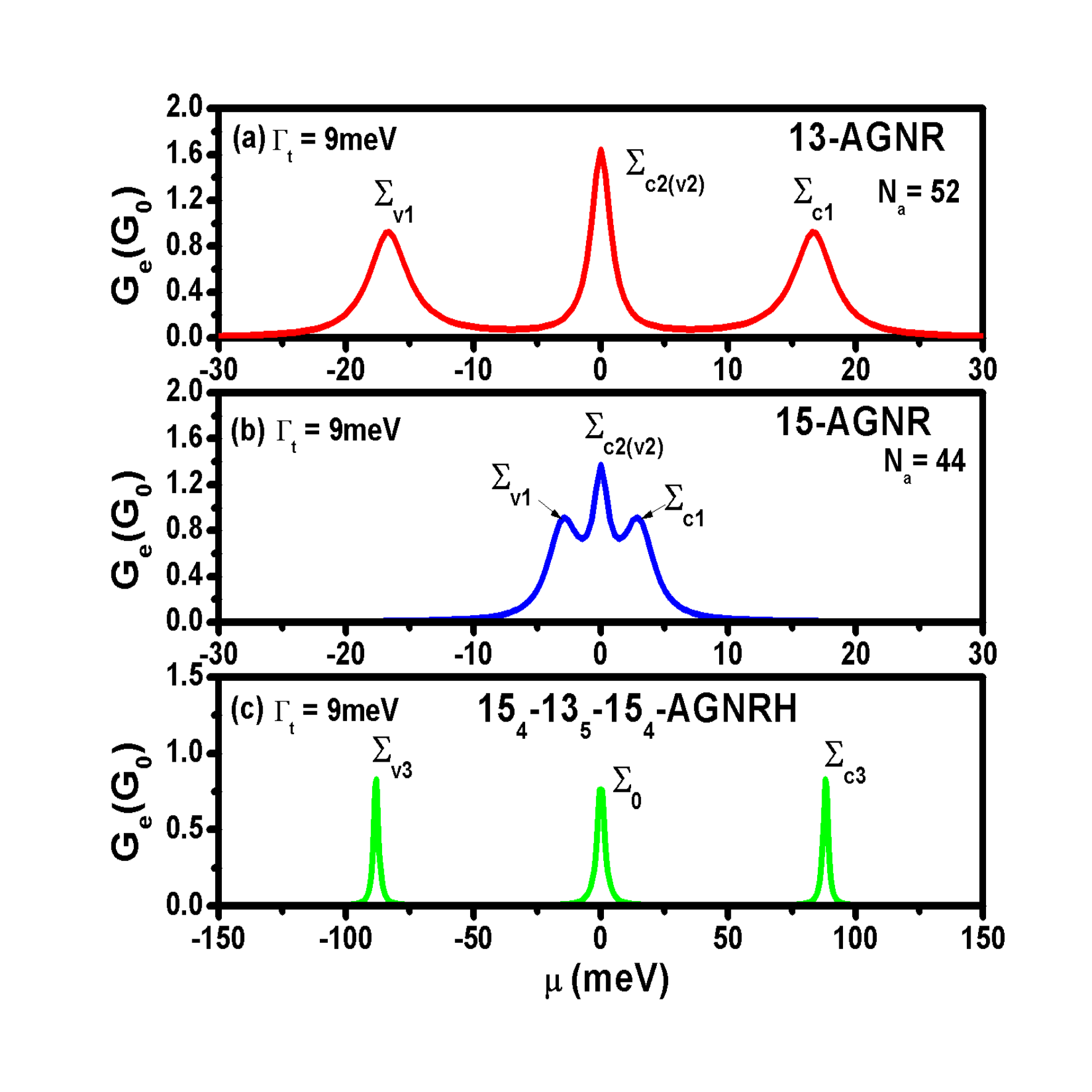}
\caption{Electrical conductance $G_e$ of GNRs with armchair edge
sites coupled to the electrodes. (a) $G_e$ of a 13-AGNR with $N_a
= 52$ as a function of $\mu$; (b) $G_e$ of a 15-AGNR with $N_a =
44$ as a function of $\mu$; (c) $G_e$ of a $15_4$-$13_5$-$15_4$
AGNRH with $N_a = 52$ as a function of $\mu$. The tunneling rate
is set to $\Gamma_t = 9$~meV.}
\end{figure}

\subsection{Tunneling Current through the Interface States of 15-13-15 AGNRHs}

Figure~6 shows that the interface states serve as the sole charge
transport channels near the CNP ($\mu = 0$) under
quasi-equilibrium conditions. We now turn to the investigation of
charge transport through the interface states of 15-13-15 AGNRHs
under nonequilibrium conditions. The multiple end zigzag edge
states can be effectively eliminated by chemical termination of
the zigzag edge sites of 15-AGNRs~[\onlinecite{GaoQ}] or by
employing graphene electrodes. In this case, the energy levels
corresponding to $\Sigma_{v3}$ and $\Sigma_{c3}$ serve as the
highest occupied molecular orbital and the lowest unoccupied
molecular orbital, respectively.

As previously mentioned, spin quantum states with longer coherence
times are desirable. Notably, two-electron spin quantum states
exhibit essential entanglement behavior~[\onlinecite{Falicov}], a
characteristic that originates from electron-electron Coulomb
interactions. Therefore, it is crucial to clarify the effect of
segment length in $15_w$-$13_x$-$15_y$ AGNRHs on the electron
Coulomb interactions and electron hopping strengths between the
interface states. The intra- and inter-interface Coulomb
interactions ($U_0$ and $U_1$) are evaluated using the expression
$\frac{1}{4\pi \kappa}\sum_{i,j}|\Psi_{L(R),I}(\textbf{r}_i)|^2
|\Psi_{L(R),I}(\textbf{r}_j)|^2\frac{1}{|\textbf{r}_i-\textbf{r}_j|},$
where $\kappa = 4$ is the dielectric constant and $U_{cc} = 4$~eV
is the on-site Coulomb interaction corresponding to double
occupancy in each $p_z$ orbital. The wave functions
$\Psi_{L(R),I}(\textbf{r}_j)$, associated with the left and right
interface states, are defined over regions $w+2 \geq j > 0$ and
$(w+x+y) \geq j \geq (w+x-2)$, respectively, thus forming the left
and right quantum dots (QDs). This definition ensures that the
charge density outside each QD region accounts for less than $5\%$
of the total charge density for $ x \ge (y+w)$.

In Fig.~8, we present the intra-dot Coulomb interaction $U_0$, the
inter-dot Coulomb interaction $U_1$, and the electron hopping
strength $t_x$ between QDs as functions of the length $x$ of the
13-AGNR segment in $15_4$-$13_x$-$15_4$ AGNRHs. As shown in
Fig.~8(a), $U_0$ saturates to a constant value when $x \geq 15$.
In contrast, the inter-dot Coulomb interaction $U_1$ decreases
with increasing $x$, as depicted in Fig.~8(b). Figure~8(c) reveals
that $t_x$ exhibits exponential decay as $x$ increases. For
comparison, we also plot $t_x$ for the $9_4$-$7_x$-$9_4$ AGNRHs.
For example, $t_x = 0.92$~meV for a $9_4$-$7_{12}$-$9_4$ AGNRH
(corresponding to a 7-AGNR segment length of 5.1~nm), while $t_x =
1.19$~meV for a $15_4$-$13_{20}$-$15_4$ AGNRH (corresponding to a
13-AGNR segment length of 8.52~nm). To quantitatively determine
the decay lengths, we fit the curves in Fig. 8(c) using the
function $e^{-x/\beta}$. The extracted decay lengths are $\beta =
3.37$ for the 15-13-15 AGNRH and $\beta = 1.86$ for the 9-7-9
AGNRH. This indicates that the decay length in the 15-13-15
structure is nearly twice as long as that in the 9-7-9 AGNRH. This
comparison indicates that it is easier to implement double gate
electrodes for controlling the interface states in 15-13-15 AGNRHs
with appropriately designed physical parameters. It is worth
noting that the physical parameters shown in Fig. 8 may not be as
precise as those obtained from first-principles calculations (DFT)
[\onlinecite{DJRizzo}]. Nevertheless, the trends in key physical
parameters--such as electron hopping strength and
electron-electron Coulomb interactions-- are consistent with DFT
results [\onlinecite{Mangnus}]. Moreover, the tight-binding model
serves as an effective tool for gaining insight into the
electronic phases of GNR heterojunctions, as predicted by DFT
methods
[\onlinecite{DJRizzo},\onlinecite{PizzocheroM},\onlinecite{TepliakovNV}].

\begin{figure}[h]
\centering
\includegraphics[angle=0,scale=0.3]{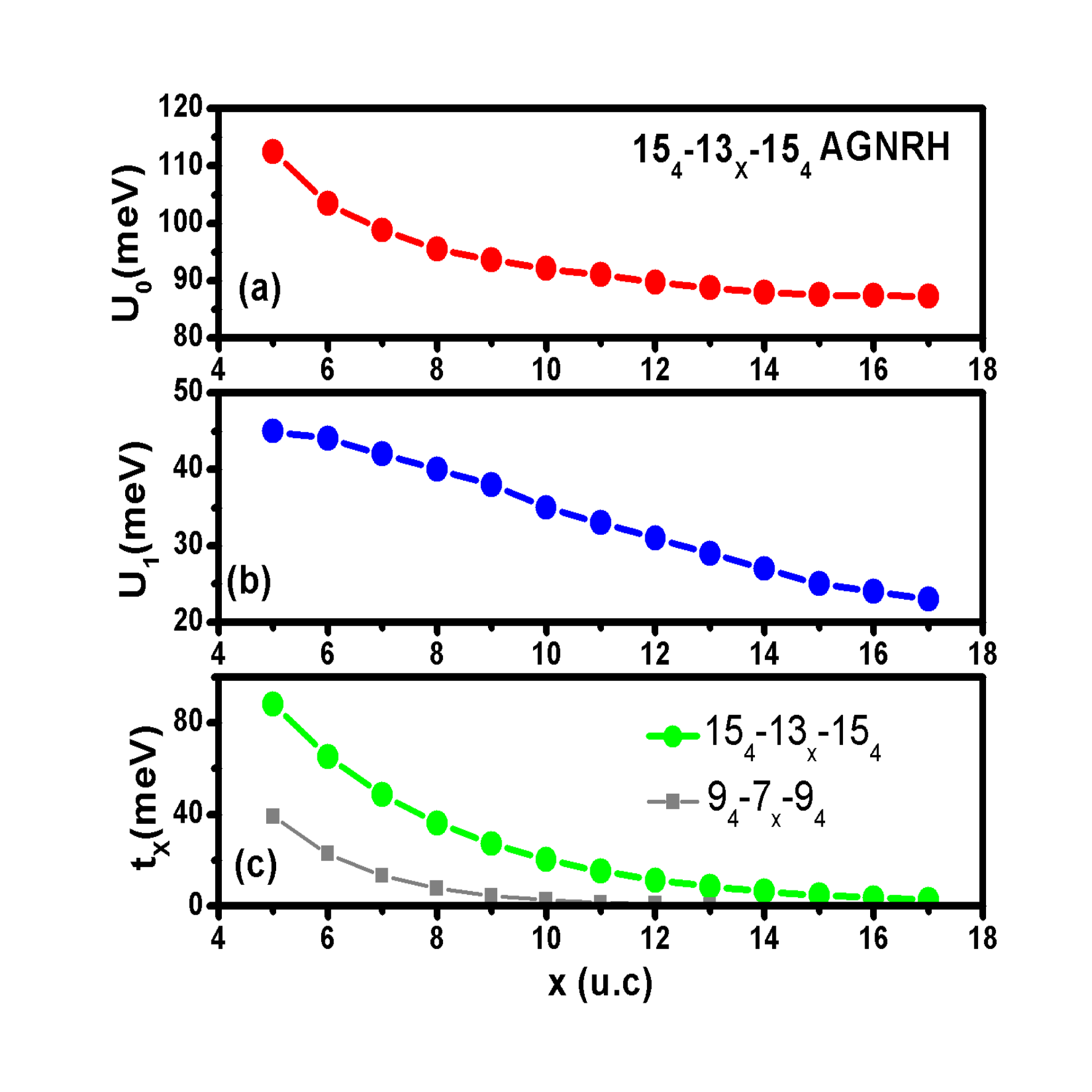}
\caption{(a) Electron Coulomb interaction of the left (right)
quantum dot, $U_0$; (b) electron Coulomb interaction between the
left and right quantum dots, $U_1$; and (c) electron hopping
strength, $t_x$, between the left and right quantum dots (QDs) of
$15_4$-$13_x$-$15_4$ and $9_4$-$7_x$-$9_4$ AGNRHs as functions of
$x$.}
\end{figure}

We now consider the extended Anderson model, including both
intra-dot and inter-dot Coulomb interactions, to study the
tunneling current through topologically protected SDQDs
illustrated in Fig. 1(c). The effective Hamiltonian of the system
is given by $H_{\text{eff}} = H_{\text{SD}} + H_{\text{SDQD}}$,
where $H_{\text{SD}}$ describes the source and drain electrodes,
and $H_{\text{SDQD}}$ represents the SDQD, expressed as

\begin{small}
\begin{eqnarray}
& &H_{SDQD}\\ \nonumber &= &\sum_{j=L,R,\sigma}E_{j}
c^{\dagger}_{j,\sigma}c_{j,\sigma}-t_{x} (c^{\dagger}_{R,\sigma}
c_{L,\sigma} + c^{\dagger}_{L,\sigma} c_{R,\sigma})\\ \nonumber
&+&\sum_{j=L,R}U_j~n_{j,\sigma}n_{j,-\sigma}
+\frac{1}{2}\sum_{j\neq\ell,\sigma,\sigma'}U_{j,\ell}~n_{j,\sigma}n_{\ell,\sigma'},
\end{eqnarray}
\end{small}
where $E_j$ is the spin-independent energy level of the $j$-th QD.
The parameters $U_j = U_{L(R)} = U_0$ and $U_{j,\ell} = U_{LR} =
U_1$ denote the intra-dot and inter-dot Coulomb interactions,
respectively. Here, $n_{j,\sigma} =
c^{\dagger}_{j,\sigma}c_{j,\sigma}$ is the number operator, and
$t_x$ is the electron hopping strength between the QDs. The
operator $c^{\dagger}_{j,\sigma}$ ($c_{j,\sigma}$) creates
(annihilates) an electron with spin $\sigma$ at site $j$. Because
the energy levels of the interface states in 15-13-15 AGNRHs are
well separated from the conduction and valence subbands by a large
bandgap (greater than 1~eV), thermal noise effects such as
phonon-assisted excitations can be neglected in the Hamiltonian of
Eq.(5).

While many theoretical studies have explored the tunneling current
in SDQDs~[\onlinecite{FranssonJ}--\onlinecite{Kondo}], relatively
few works have addressed the effects of temperature on current
rectification under the Pauli spin blockade regime. In this study,
we employ the Keldysh-Green function
technique~[\onlinecite{Kuo5},\onlinecite{Kuo3},\onlinecite{LandiGT}]
to derive the tunneling current through an SDQD coupled to
metallic electrodes. The tunneling current leaving the left
(right) electrode is given by

\begin{equation}
J_{L(R)}(V_{SD},T)=\frac{2e}{h}\int {d\varepsilon}~ {\cal
T}_{LR(RL)}(\varepsilon)[f_L(\varepsilon)-f_R(\varepsilon)]
\end{equation}
where
$f_{\alpha}(\varepsilon)=1/\{\exp[(\varepsilon-\mu_{\alpha})/k_BT]+1\}$
is the Fermi distribution function for the $\alpha$-th electrode.
The chemical potentials are defined as $\mu_{L(R)} = \mu \pm
eV_{SD}/2$, with $\mu = 0$ set at the Fermi energy. Here,
$\mathcal{T}_{LR}(\varepsilon)$ represents the transmission
coefficient of the SDQD system, and it has a closed-form
expression (see Appendix A). To obtain the reversed-bias tunneling
current, it suffices to exchange the indices of
$\mathcal{T}_{LR}(\varepsilon)$ in Eq.(A.1). It is important to
note that the expression in Eq.(A.1) is valid only in the Coulomb
blockade regime and not in the Kondo
regime~[\onlinecite{JeongH}--\onlinecite{WangJN}].

\begin{figure}[h]
\centering
\includegraphics[angle=0,scale=0.3]{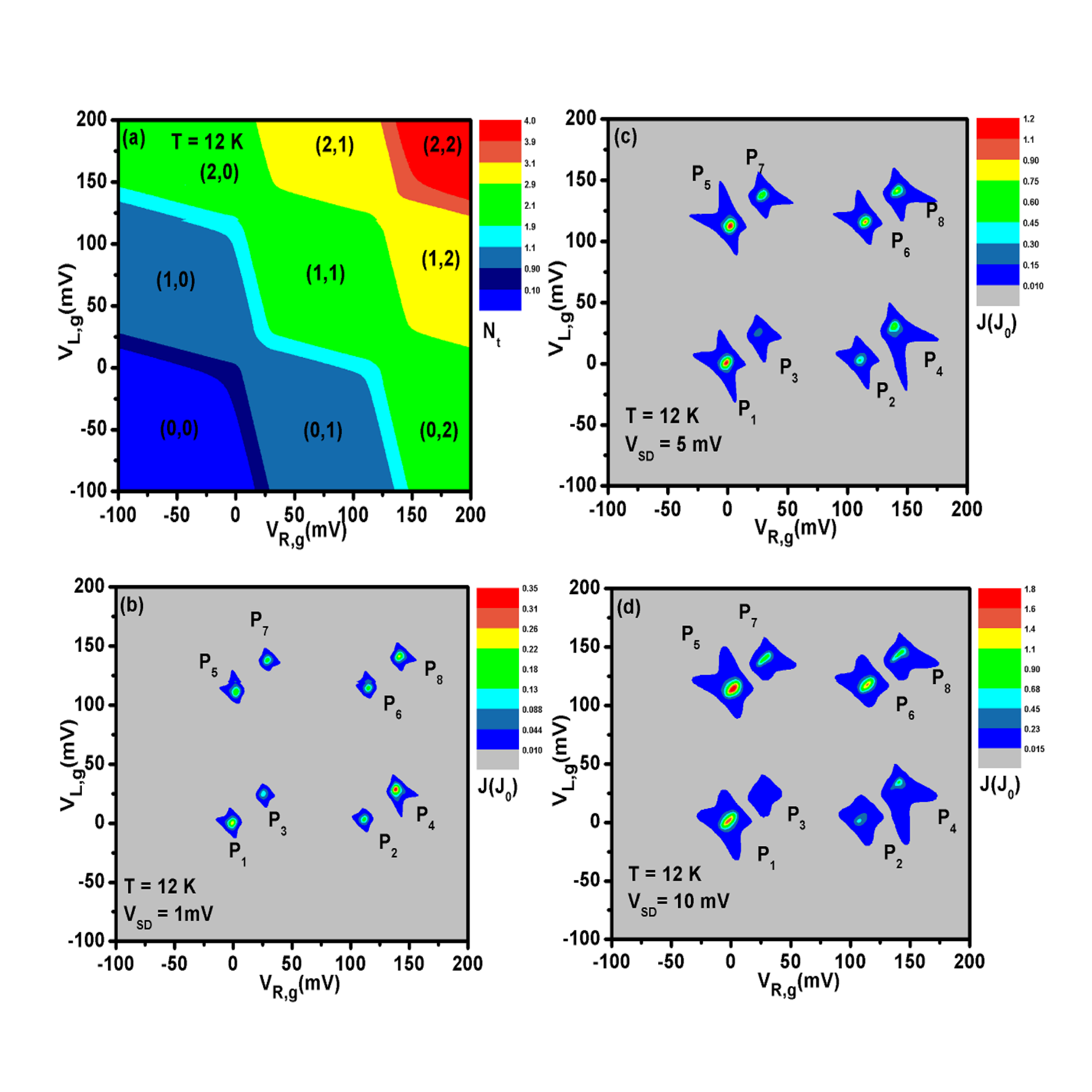}
\caption{(a) Total occupation number, $N_t =
\sum_{\sigma}(N_{L,\sigma} + N_{R,\sigma})$, as a function of gate
voltages $V_{L,g}$ and $V_{R,g}$ at a temperature of $T = 12$~K
and a small applied voltage $V_{SD} = 1$~mV. (b) Tunneling current
$J_L$ at $V_{SD} = 1$~mV, (c) tunneling current $J_L$ at $V_{SD} =
5$~mV, and (d) tunneling current $J_L$ at $V_{SD} = 10$~mV as
functions of gate voltages $V_{L,g}$ and $V_{R,g}$ at a
temperature of $T = 12$~K. The following physical parameters were
adopted: $U_0 = 90$~meV, $U_1 = 25$~meV, $t_x = 1$~meV, $\eta =
0.1 $ and $\Gamma_{e,L} = \Gamma_{e,R} = \Gamma_{e,t} = 1$~meV.
The tunneling current magnitude is described by $J$ with $V_{SD} >
0$, and $\delta_1 = \delta_2 = 0.8$ was considered. Tunneling
currents are expressed in units of $J_0 = 0.773$~nA.}
\end{figure}

Based on Eq.~(6) and Eq. (A.1), we present the charge transport
characteristics of the SDQD under various applied voltages in
Fig.9. Figure 9(a) displays the charge stability diagram as
functions of the two gate voltages, $V_{L,g}$ and $V_{R,g}$, at a
small bias voltage $V_{SD} = 1$~mV and temperature $T = 12$~K.
Each plateau labeled by $(N_L, N_R)$ represents a specific charge
configuration of the SDQD. For example, $(0,0)$ corresponds to an
empty SDQD, where both energy levels, $E_L$ and $E_R$, lie above
the Fermi level ($\mu = 0$). Four narrow lines are observed,
corresponding to regions where particle-number transitions occur.
The widths of these transition regions are determined by thermal
energy ($k_BT$), contact broadening ($\Gamma_{e,t}$), electron
hopping strength ($t_x$) and the applied bias voltage ($V_{SD}$).
The honeycomb-shaped plateau regions reflect the strength of
electron-electron Coulomb interactions. The results shown in Fig.
9(a) are consistent with those reported in
Refs.[\onlinecite{DasSarma},\onlinecite{DasSarma1}].

Figure~9(b) shows the tunneling current at $V_{SD} = 1$mV. Current
spectra featuring eight distinct peaks appear at the
particle-number transition regions, corresponding to the eight
transport configurations described in Eq.(A.1). These current
peaks reveal that charge transport occurs only within narrow
regions defined by the two gate voltages, $V_{L,g}$ and $V_{R,g}$.
As the applied bias $V_{SD}$ increases, the charge transport
regions broaden, as illustrated in Figs. 9(c) and 9(d). Such
bias-dependent expansion of the transport regions has also been
observed experimentally in different types of SDQD
systems[\onlinecite{vanderWiel},\onlinecite{JphnsonAC},\onlinecite{BorselliMG}].
To the best of our knowledge, charge stability diagrams for GNR
nanostructures have not yet been experimentally reported, as
fabricating double gate electrodes for 7-AGNR and 9-AGNR segments
remains challenging [\onlinecite{ZhangJ}--\onlinecite{ZhangJain}].
Notably, the transport configurations labeled by $P_2$ and $P_5$
exhibit pronounced current rectification behavior, which arises
from many-body effects~[\onlinecite{Ono}].

\subsection{Room Temperature Current Rectification in the Pauli Spin Blockade Configuration}

To further explore the nonlinear behavior of charge transport, we
present the tunneling current as functions of $V_g$ and $V_{SD}$
at $T = 12$~K, with $V_{L,g} = V_{R,g} = V_g$ (functioning as a
single gate electrode), in Fig.~10(a). Under the condition
$V_{L,g} = V_{R,g}$, the magnitudes of the tunneling currents
satisfy $J_L = |J_R| $, where $J_R$ is the electron current
leaving the right electrode for $V_{SD} < 0$. This symmetry
indicates the absence of current rectification, although negative
differential conductance (NDC) behavior emerges at higher applied
voltages. The color-contour plot in Fig.10(a) reveals several
diamond-shaped regions arising from electron Coulomb interactions
[\onlinecite{ZhangJ}--\onlinecite{ZhangJain}].

For charge transport in the Pauli spin blockade (PSB)
configuration, achieved by setting $eV_{L,g} = |4U_1-U_0|/3$ and
$eV_{R,g} = |4U_0-U_1|/3$ (corresponding to the $P_2$ peak in
Fig.~9(b)), we present the tunneling current as functions of
$V_{SD}$ and $T$ in Fig.~10(b). Due to the resonant channel
condition $E_L + U_1 = E_R + U_0 = \mu$, where $\mu$ is the Fermi
energy, $J$ shows a linear dependence on $V_{SD}$ at small bias
voltages. At larger $V_{SD}$, significant current rectification is
observed at $T = 12$~K. The maximum backward tunneling current
$J_{max,B} = 0.99$~nA ($V_{SD} < 0$) is much larger than the
maximum forward current $J_{max,F} = 0.318$~nA ($V_{SD}
> 0$). This type of current rectification has previously been
reported experimentally in GaAs SDQDs at extremely low
temperatures[\onlinecite{Ono}]. Extensive theoretical and
experimental studies have focused on this current rectification
phenomenon at near-zero
temperatures~[\onlinecite{FranssonJ}--\onlinecite{Kuo3}]. The
suppression of forward tunneling currents is attributed to the
occupation of two-electron triplet states in the
SDQD~[\onlinecite{FranssonJ}--\onlinecite{Inarrea},\onlinecite{Kuo3}].
However, as the temperature increases, the current rectification
effect disappears for $k_BT \geq 13$~meV ($156$~K).

Next, we consider the case of asymmetric tunneling rates, which
can be realized by introducing asymmetrical 15-AGNR segment
lengths in a 15-13-15 AGNRH. At $k_BT = 20$~meV, current
rectification is still observed, as shown in Fig.~10(c) with
$\Gamma_{e,L} = 3$~meV and $\Gamma_{e,R} = 1$~meV. In Fig.~10(d),
we consider the opposite asymmetry with $\Gamma_{e,L} = 1$~meV and
$\Gamma_{e,R} = 3$~meV. Comparing Figs.10(c) and 10(d), we find
that the rectification spectra differ significantly. This
observation demonstrates that current rectification in the PSB
configuration originates from many-body interactions rather than
contact properties related to the electrodes and the SDQD (or
molecule)[\onlinecite{TaoNJ},\onlinecite{AndrewsDQ}].

\begin{figure}[h]
\centering
\includegraphics[angle=0,scale=0.3]{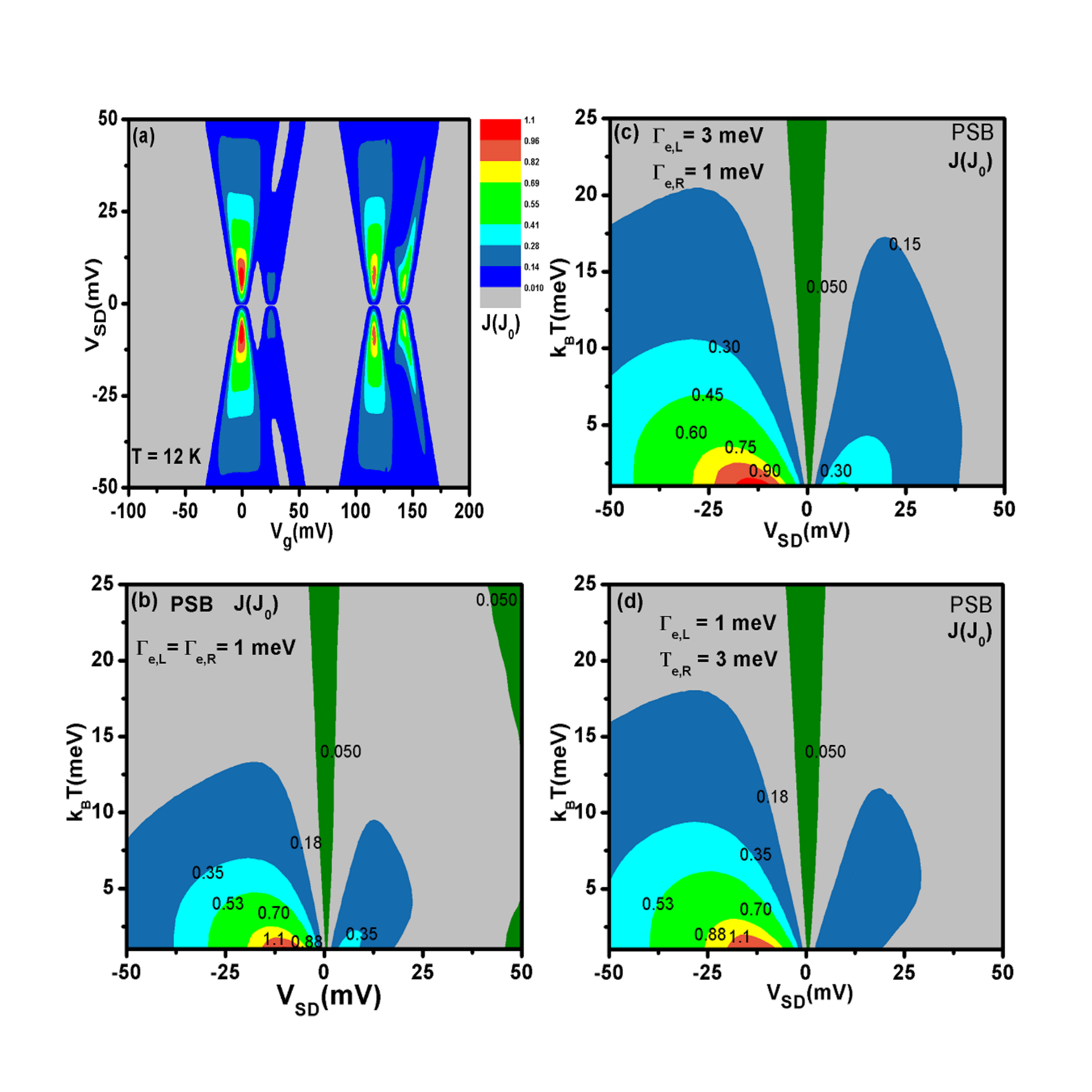}
\caption{(a) Tunneling current as a function of $V_{SD}$ and $V_g$
at $T = 12$~K, with $\Gamma_{e,L} = \Gamma_{e,R} = 1$~meV.
Tunneling current as a function of $V_{SD}$ and $k_B T$ in the
Pauli spin blockade configuration, where $eV_{L,g} = |4U_1 -
U_0|/3$ and $eV_{R,g} = |4U_0 - U_1|/3$. (b) $\Gamma_{e,L} =
\Gamma_{e,R} = 1$~meV, (c) $\Gamma_{e,L} = 3~meV$ and
$\Gamma_{e,R} = 1$~meV, and (d) $\Gamma_{e,L} = 1$~meV and
$\Gamma_{e,R} = 3$~meV. Other physical parameters are the same as
those in Figure 9.}
\end{figure}

Although significant current rectification in SDQDs can still
occur at elevated temperatures by enlarging one of the tunneling
rates, the ratio $J_{max,B}/J_{max,F}$ in Fig.10 approaches unity
at room temperature. Recently, researchers have raised the
question: "Is it possible to observe PSB current rectification at
room temperature?"[\onlinecite{BanY}]. To investigate this
intriguing phenomenon, we analyze the tunneling current described
by Eq.~(6).

The temperature dependence of the tunneling current in Eq.~(6) is
influenced by two primary factors: the Fermi distribution function
of the electrodes and the transmission coefficient of the SDQD.
When the carrier concentration in the electrodes is not affected
by doping, the Fermi energy $\mu$ remains temperature-independent.
Thus, while the Fermi distribution function exhibits typical
thermal broadening with temperature, $\mu$ remains fixed in our
study. Meanwhile, the temperature dependence of the transmission
coefficient arises from many-body effects, which modify the system
correlation functions.

To better understand how temperature impacts current
rectification, we present in Fig.~11 the calculated particle
occupation numbers, interdot two-particle correlation functions
for triplet and singlet states, and the tunneling current in the
PSB configuration without orbital offset ($\eta = 0$). In
Fig.~11(a), the total particle occupation number $N_t$ remains
within the range $1.5 \leq N_t \leq 2$ across a wide temperature
range ($T = 12$~K to $T = 300$~K). At $T = 12$~K and $V_{SD} > 0$,
the single-particle occupation numbers $N_{L,\sigma}$ and
$N_{R,\sigma'}$ both approach one-half, as shown in Figs.~11(b)
and 11(c), respectively. When $V_{SD} < 0$ at $T = 12$~K,
$N_{L,\sigma}$ and $N_{R,\sigma'}$ tend toward $N_{L,\sigma} =
0.1$ and $N_{R,\sigma'} = 1$, respectively. It is important to
note that $N_{L,\sigma}$ does not vanish completely because the
right electrode can still inject particles into the left dot via
the formation of a two-particle spin singlet state of the right
dot.

As temperature increases, $N_{L,\sigma}$ decreases for $V_{SD} >
0$ but increases for $V_{SD} < 0$, while $N_{R,\sigma'}$
consistently decreases with increasing temperature. At low
temperature ($T = 12$~K), the results of Figs.~11(d) and 11(e)
indicate that the SDQD is predominantly occupied by a inter-dot
two-particle triplet state when $V_{SD} > 0$. However,
distinguishing between different inter-dot two-particle spin
states becomes increasingly difficult at higher temperatures. The
results in Fig.~11(f) show that current rectification in the PSB
configuration is less sensitive to temperature variation when
there is no orbital offset ($\eta = 0$), compared to the results
in Fig.~10(b). According to the probability weight of the $P_2$
configuration, minimizing the temperature effect on
$N_{R,\sigma}$, $\langle n_{R,\sigma} n_{L,\sigma} \rangle$, and
$\langle n_{-R,\sigma} n_{R,\sigma} \rangle$ is crucial--
particularly for $N_{R,\sigma}$ and $\langle n_{R,\sigma}
n_{L,\sigma} \rangle$. This suggests the need for a deeply bound
$E_R$ level, located well below $\mu_R$, while satisfying the
resonance condition $\epsilon_L (E_L + U_1) = \epsilon_R (E_R +
U_0) = \mu$. For $\eta = 0$, the energy levels $\epsilon_L$ and
$\epsilon_R$ can be positioned far from the chemical potentials
$\mu_L = \mu + eV_{SD}/2$ and $\mu_R = \mu - eV_{SD}/2$,
respectively, thereby reducing the sensitivity of the transport
properties to variations in the Fermi distribution function.

\begin{figure}[h]
\centering
\includegraphics[angle=0,scale=0.3]{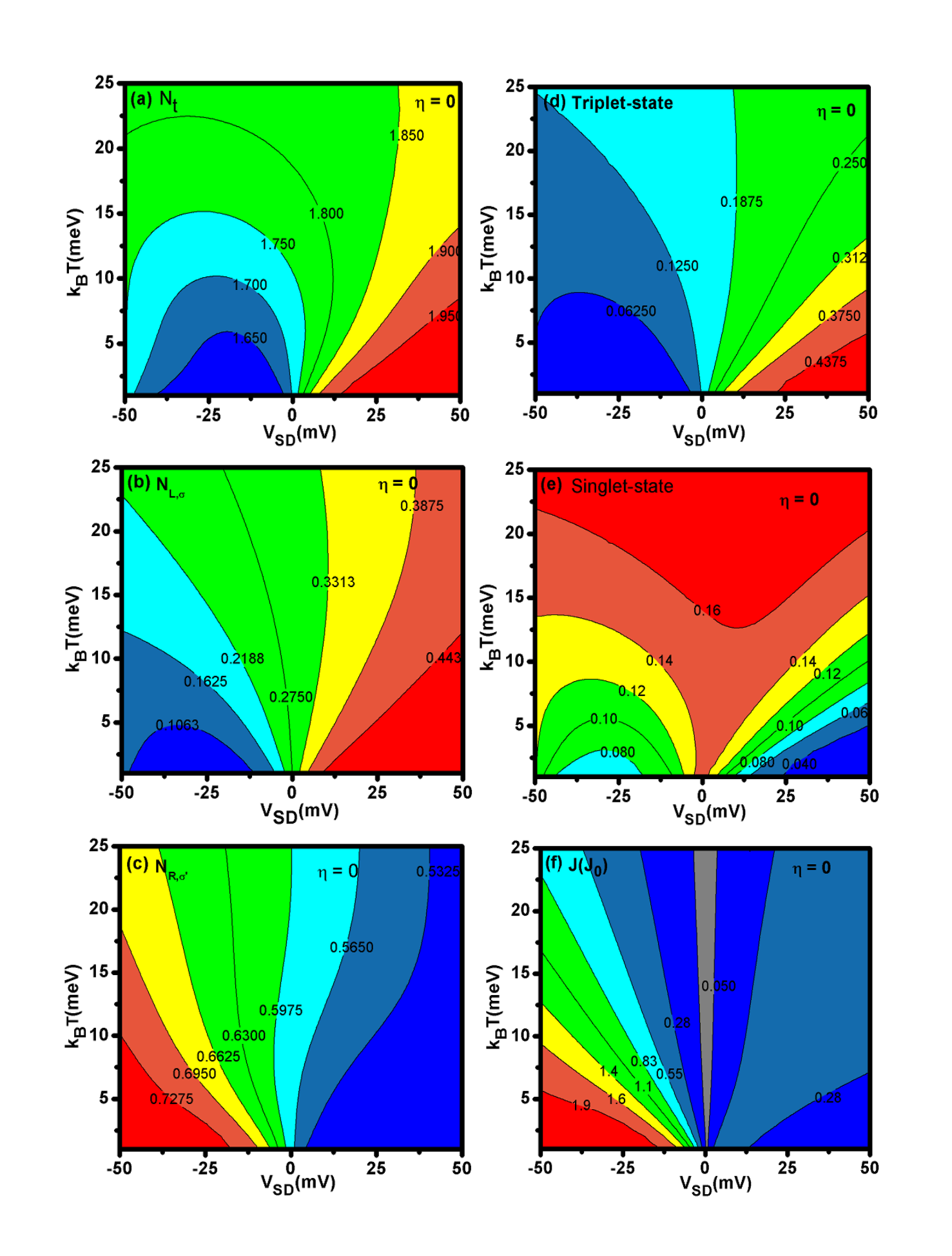}
\caption{(a) Total particle occupation number, $N_t$; (b)
single-particle occupation number of the left dot with spin
$\sigma$, $N_{L,\sigma}$; (c) single-particle occupation number of
the right dot with spin $\sigma'$, $N_{R,\sigma'}$; (d) inter-dot
two-particle correlation function for the triplet state; (e)
inter-dot two-particle correlation function for the singlet state;
and (f) tunneling current ($J$) as functions of $V_{SD}$ and $T$
in the Pauli spin blockade (PSB) configuration with $\eta = 0$.
Other physical parameters are the same as those in Figure 10(b).}
\end{figure}

As illustrated in Fig.~11, maintaining the resonant channel
$\epsilon_L = \epsilon_R = \mu$ is crucial for achieving
significant current rectification. To suppress the effects of
orbital offset, it is necessary to relax the constraint of weak
coupling between the electrodes and the SDQD (as shown in Fig.
10(c) and 10(d)). In Fig.~12, we present the tunneling current of
the SDQD under the PSB configuration as a function of the applied
voltage $V_{SD}$ and the electron hopping strength $t_x$, while
maintaining the tunneling rate $\Gamma_{e,t} = t_x$ to optimize
the tunneling current.

In Figs.~12(a) and 12(b), we consider a weak orbital offset ($\eta
= 0.1$) and show the tunneling current at $T = 12$~K and $T =
300$~K, respectively. On the other hand, a strong orbital offset
($\eta = 0.3$) is considered in Fig.~12(c) at $T = 12$~K and in
Fig.~12(d) at $T = 300$~K. As seen in Fig.12(a), under low
temperature and weak orbital offset, current rectification is
observed over a wide range of $t_x$. In particular, the tunneling
current is enhanced by nearly an order of magnitude when $t_x =
\Gamma_{e,t}$ is increased from 1~meV to 10~meV. Moreover, as $t_x
= \Gamma_{e,t}$ increases, the maximum forward and backward
tunneling currents ($J_{max,F}$ and $J_{max,B}$) shift toward
higher applied voltages, indicating that the region of NDC also
moves to higher $V_{SD}$. At $T = 300$~K, as shown in Fig.~12(b),
current rectification is suppressed for $t_x = \Gamma_{e,t} =
1$~meV. However, significant current rectification persists at
$t_x = \Gamma_{e,t} = 10$~meV.  We find that the ratios
$J_{max,B}/J_{max,F}$ are 2.12 and 1.95 for $T = 12$~K and $T =
300$~K, respectively, in the case of $t_x = \Gamma_{e,t} =
10$~meV.

For strong orbital offset ($\eta = 0.3$), the ratio
$J_{max,B1}/J_{max,F}$ approaches unity even at $T = 12$~K and
$t_x = \Gamma_{e,t} = 10$~meV, as shown in Fig.~12(c). Notably, an
additional peak labeled $J_{B2}$ emerges when $t_x = \Gamma_{e,t}
\geq 3$~meV, with the $J_{B2}$ tunneling current surpassing
$J_{B1}$ at $t_x = 10$~meV. This additional current originates
from a configuration where the SDQD holds two particles in the
left dot and one particle in the right dot, characterized by the
transmission coefficient:

\begin{eqnarray}
& &{\cal T}_{RL}(P_8)/(4\Gamma_{e,L}t^2_{x}\Gamma_{e,R})\nonumber \\
&=&\frac{\langle
n_{L,-\sigma}n_{L,\sigma}n_{R,\sigma}\rangle}{|(\epsilon_L-U_L-2U_{LR})(\epsilon_R-U_R-2U_{LR})-t^2_{x}|^2}.
\end{eqnarray}

Since we set $E_L + U_{LR} = E_R + U_R = \mu = 0$ for the PSB
configuration, the tunneling associated with $J_{B2}$ occurs at an
applied voltage of $eV_{SD} = (-U_L + U_{LR})/(2\eta)$ for small
$t_x$. The probability of double occupancy in the left dot
increases as the broadening of $E_L$ grows, explaining why
$J_{B2}$ strengthens with increasing $t_x = \Gamma_{e,t}$. As
shown in Fig.~12(d), $J_{B2}$ is significantly enhanced at $T =
300$~K even for a small $t_x = \Gamma_{e,t} = 1$~meV, due to
thermally assisted two-particle occupation in the left dot.
Meanwhile, at $T = 300$~K, the original current rectification
between $J_F$ and $J_{B1}$ is suppressed. Instead, rectification
between $J_{B2}$ and $J_F$ becomes dominant under strong orbital
offset conditions.

\begin{figure}[h]
\centering
\includegraphics[angle=0,scale=0.3]{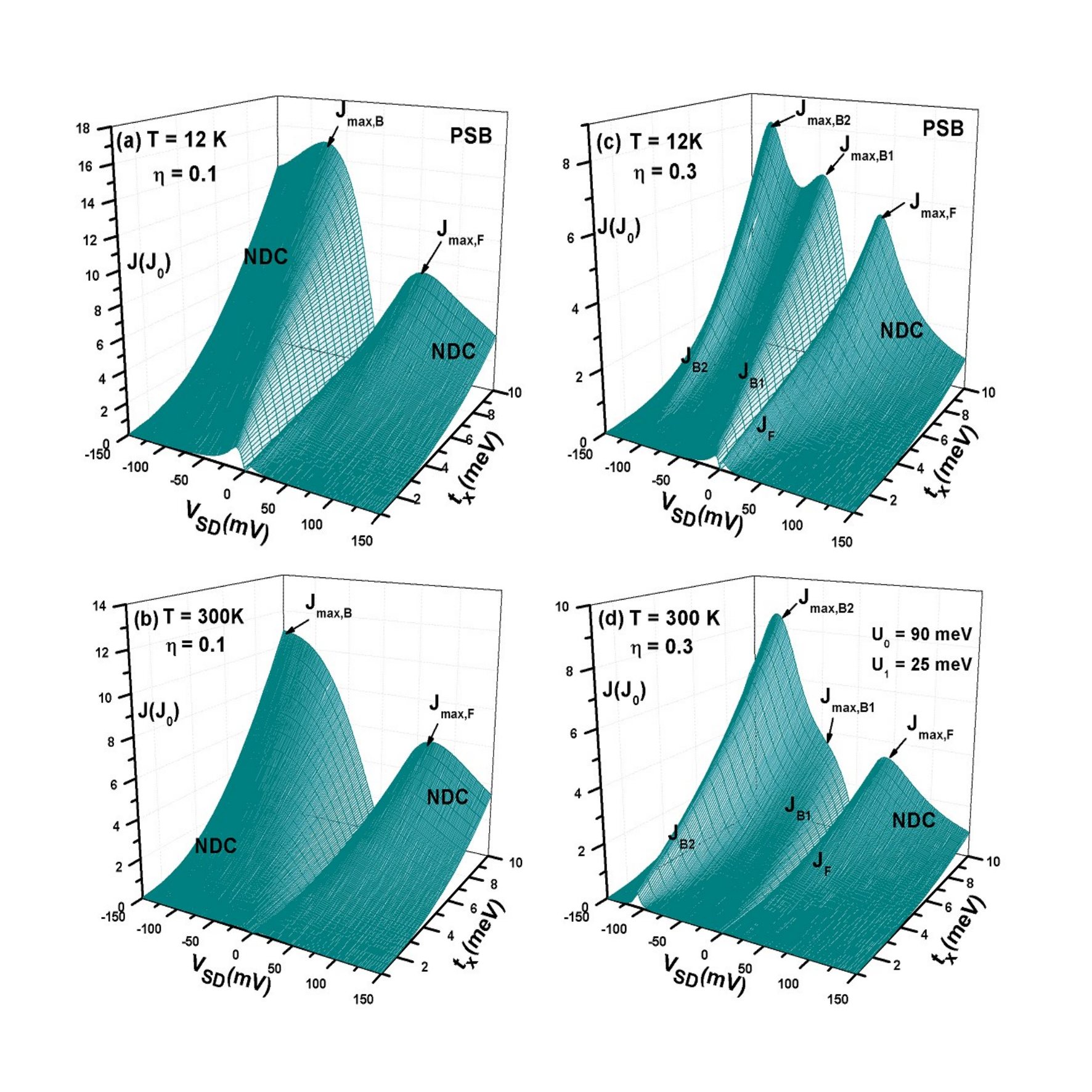}
\caption{Tunneling current of the SDQD in the Pauli spin blockade
(PSB) configuration as a function of $V_{SD}$ and $t_x$ under the
condition $t_x = \Gamma_{e,t}$, $U_0 = 90$~meV, and $U_1 =
25$~meV. (a) $T = 12$~K and $\eta = 0.1$, (b) $T = 300$~K and
$\eta = 0.1$, (c) $T = 12$~K and $\eta = 0.3$, and (d) $T = 300$~K
and $\eta = 0.3$.}
\end{figure}

Because the current rectification of the SDQD in the PSB
configuration is crucial for applications such as spin-current
conversion [\onlinecite{Ono}], it is important to evaluate the
two-particle correlation functions of inter-dot triplet and
singlet states along the trajectory of the maximum forward
current, $J_{max,F}$. In Fig.~13, we present the calculated
two-particle correlation functions of inter-dot triplet and
singlet states as functions of $V_{SD}$ and $t_x$ at $\eta = 0.1$.
From these results, we extract the temperature-dependent
probabilities of the triplet and singlet states. At $T = 12$~K,
these two correlation functions exhibit a significant difference
along the $J_{max,F}$ trajectory, with the triplet state
probability being approximately twice that of the singlet state.
However, at $T = 300$~K, the difference between the triplet and
singlet correlations becomes much smaller.

Although the ratio $J_{max,B}/J_{max,F}$ at $T = 300$K can be
improved by increasing particle Coulomb interactions, we do not
present such results in this study. One possible approach to
enhance intra-dot and inter-dot Coulomb interactions is to
introduce vacancies into the 15-AGNR segments of 15-13-15 AGNRHs,
which can be done without disrupting the interface states of
AGNRHs [\onlinecite{Kuo7}].  While enhancing Coulomb interactions
typically requires smaller QD sizes, this poses challenges for the
layout of double-gate electrodes [\onlinecite{BanY},
\onlinecite{FrancescoR}]. In Ref. [\onlinecite{BanY}], the authors
investigated a SDQD system composed of atomic-scale quantum dots
formed by sulfur ($S$) and zinc ($Zn$) atoms, which act as the
left and right QDs, respectively. However, due to the random
distribution associated with doping, it is difficult to precisely
control the positions of the $S$ and $Zn$ atoms. Consequently, it
becomes challenging to accurately define and tune the double gate
voltages ($V_{L,g}$ and $V_{R,g}$) needed to manipulate the
resonant tunneling conditions required for Pauli spin blockade
[\onlinecite{Zajac}].

\begin{figure}[h]
\centering
\includegraphics[angle=0,scale=0.3]{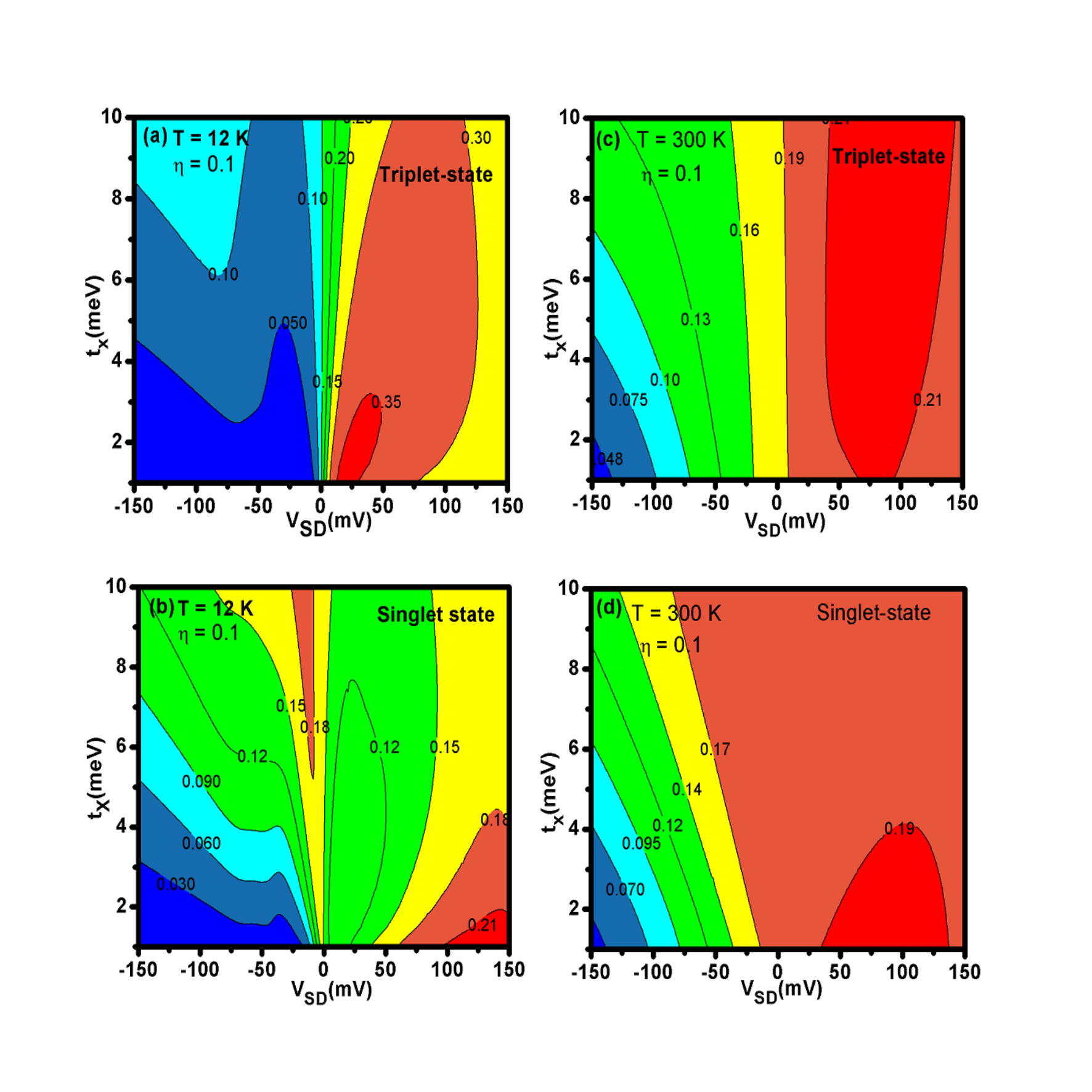}
\caption{Two-particle correlation function of inter-dot triplet
and singlet states as a function of $V_{SD}$ and $t_x$ at weak
orbital offset $\eta = 0.1$. (a) Triplet state at $T = 12$~K, (b)
Singlet state at $T = 12$~K, (c) Triplet state at $T = 300$~K, and
(d) Singlet state at $T = 300$~K. Other physical parameters are
the same as those in Figure 12. Note that $t_x = \Gamma_{e,t}$.}
\end{figure}

\section{Conclusion}
In this work, we investigated the electronic structures of
13-11-13 and 15-13-15 AGSLs using a tight-binding model. We found
that the conduction and valence subbands of 15-13-15 AGSLs can be
accurately described by the SSH model, where topologically
protected interface states emerge at the junctions between 15-AGNR
and 13-AGNR segments. This indicates that finite-length 15-13-15
AGSLs act as QD arrays [\onlinecite{LuoYi}], where each dot hosts
a single discrete energy level well separated from the bulk states
by an energy gap significantly larger than the thermal energy at
room temperature (see Figs. 2(c) and 2(d)). Such characteristics
highlight the potential of these systems for use in solid-state
quantum processors operable at elevated temperatures
[\onlinecite{DiVincenzo}, \onlinecite{LloydS}].

For 15-13-15 AGNRH segments, both multiple zigzag edge states at
the ends and topologically protected interface states are observed
under longitudinal electric fields. Owing to their spatial
localization, these states exhibit linear Stark shifts in their
energy levels. However, the highly localized wave-functions of the
end zigzag edge states hinder efficient electronic transport when
coupled to electrodes. In contrast, the interface states provide
robust tunneling pathways. The conductance formula derived in Eq.
(4) demonstrates that these states behave effectively as a SDQD
system. Although similar features are present in 9-7-9 AGNRH
segments, the electron hopping strength ($t_x$) in 15-13-15 AGNRH
exhibits nearly twice the decay length compared to that in 9-7-9
AGNRH, as shown in Fig. 8(c). Consequently, the 15-13-15 AGNRH
segment provides a size advantage for implementing two-gate
electrode configurations.

We further explored nonlinear charge transport through these
topologically protected interface states under PSB, employing a
two-site Anderson model that incorporates both intra-dot and
inter-dot Coulomb interactions. The analysis, carried out using
the Keldysh Green function formalism, reveals tunneling current
spectra exhibiting charge stability diagrams and Coulomb blockade
oscillations, consistent with experimental results in other SDQD
systems [\onlinecite{Ono}, \onlinecite{JphnsonAC},
\onlinecite{BorselliMG}]. Due to the extremely short decay lengths
of the end zigzag states, few experimental studies have reported
charge stability diagrams originating from topological states in
7-AGNRs and 9-AGNRs. In contrast, single-electron tunneling
transport has been observed in GNR QDs
[\onlinecite{ZhangJ}--\onlinecite{ZhangJain}].

The Stark effect observed in the energy levels indicates that
bias-induced orbital offsets significantly influence the tunneling
current spectra. Under weak orbital offset conditions ($\eta \leq
0.1$), strong current rectification persists across a broad
temperature range, particularly when the level broadening
satisfies $\Gamma_{e,t} = t_x$. The ratio between the maximum
backward and forward tunneling currents reaches approximately two
for $\eta = 0.1$ and $\Gamma_{e,t} = t_x = 10$~meV at $T = 300$~K.
In contrast, under strong orbital offset conditions ($\eta \geq
0.3$), the PSB-induced current rectification channel ($E_L +
U_{LR} = E_R + U_R $) is largely suppressed. Instead, a new
thermally activated tunneling channel ($E_L + U_L + 2U_{LR} = E_R
+ U_R + 2U_{RL}$) dominates, leading to substantial reversed
current at room temperature.

Our results demonstrate that robust current rectification in SDQDs
based on the interface states of 15-13-15 AGNRHs can be achieved
at temperatures above 12~K, offering a promising platform for
spin-current conversion applications [\onlinecite{Ono}]. Compared
to other structures proposed in the literature [\onlinecite{BanY},
\onlinecite{KoderaT}--\onlinecite{MaRL}], AGNRHs present several
advantages. Notably, the electronic properties of the interface
states--including electron-electron interactions, hopping
amplitudes, and tunneling rates--can be precisely tuned via
bottom-up synthesis techniques
[\onlinecite{ChenYC}--\onlinecite{SongST}]. Additionally, the
integration of two-gate electrodes is feasible using existing
semiconductor fabrication processes [\onlinecite{LlinasJP}],
further advancing the practical development of AGNR-based quantum
devices.


{}

{\bf Data Availability Statements}\\

The data that supports the findings of this study are available
within the article.

\textbf{Conflicts of interest }\\

There are no conflicts to declare.\\

{\bf Acknowledgments}\\
{This work was supported in part by the Ministry of Science and
Technology (MOST), Taiwan under Contract No. 107-2112-M-008-023MY2
}

\mbox{}\\
E-mail address: mtkuo@ee.ncu.edu.tw\\

\appendix{APPENDICES}

\numberwithin{figure}{section}

\numberwithin{equation}{section}

\section{Transmission coefficient}
According to the tunneling current expression in Eq. (6), electron
transport through SDQDs in the Coulomb blockade region is governed
by the transmission coefficient shown below:
\begin{small}
\begin{eqnarray}
& &{\cal T}_{LR}(\epsilon)/(4t^2_{x}\Gamma_{e,L}\Gamma_{e,R})=\frac{P_{1} }{|\epsilon_L\epsilon_R-t^2_{x}|^2} \nonumber \\
&+& \frac{P_{2} }{|(\epsilon_L-U_{LR})(\epsilon_R-U_R)-t^2_{x}|^2} \nonumber \\
&+& \frac{P_{3} }{|(\epsilon_L-U_{LR})(\epsilon_R-U_{LR})-t^2_{x}|^2} \label{TF} \\
\nonumber &+&
\frac{P_{4} }{|(\epsilon_L-2U_{LR})(\epsilon_R-U_{LR}-U_R)-t^2_{x}|^2}\\
\nonumber &+& \frac{P_{5} }{|(\epsilon_L-U_{L})(\epsilon_R-U_{LR})-t^2_{x}|^2}\\
\nonumber &+& \frac{P_{6}
}{|(\epsilon_L-U_L-U_{LR})(\epsilon_R-U_R-U_{LR})-t^2_{x}|^2}\\
\nonumber &+&
\frac{P_{7} }{|(\epsilon_L-U_L-U_{LR})(\epsilon_R-2U_{LR})-t^2_{x}|^2}\\
\nonumber
 &+&
\frac{P_{8} }{|(\epsilon_L-U_L-2U_{LR})(\epsilon_R-U_R-2U_{LR})-t^2_{x}|^2}, \\
\nonumber
\end{eqnarray}
\end{small}
where $\epsilon_L = \varepsilon - E_L + i\Gamma_{e,L}$ and
$\epsilon_R = \varepsilon - E_R + i\Gamma_{e,R}$. Equation~(A.1)
describes the transmission pathways for an electron with spin
$-\sigma$ in the left electrode tunneling through the SDQD via
eight distinct
configurations~[\onlinecite{Kuo5},\onlinecite{Kuo3}]. The energy
levels, modified by the applied voltage, are given by $E_L =
-(\delta_1eV_{L,g}+(1-\delta_1)eV_{R,g})+\eta~eV_{SD}$ and $E_R =
-((1-\delta_2)eV_{L,g}+\delta_2V_{R,g})-\eta~eV_{SD}$, where
$\delta_1$ and $\delta_2$ are determined by the intra- and
inter-dot Coulomb
interactions~[\onlinecite{DasSarma},\onlinecite{DasSarma1}]. The
parameter $\eta < 1/2$ adjusts for the orbital energy offset
induced by the source-drain bias $V_{SD}$ (as shown in Fig.~3(d)).
For symmetrical $15_w-13_x-15_y$ AGNRH segments with $w = y $, the
value of $\eta$ can be estimated by $\frac{x/2}{2w+x}$. For a
given 13-AGNR segment, $\eta$ can be modulated by varying the
length 15-AGNR segments. Here, $V_{L,g}$ and $V_{R,g}$ denote the
left and right gate voltages, respectively.

The probability weights $P_m$ associated with each configuration
are given by below

\begin{small}
\begin{eqnarray}
P_{1}&=&1-N_{L,\sigma}-N_{R,\sigma}-N_{R,-\sigma}+ \langle
n_{R,\sigma}n_{L,\sigma}\rangle \nonumber \\ &+&\langle
n_{R,-\sigma}n_{L,\sigma}\rangle+\langle
n_{R,-\sigma}n_{R,\sigma}\rangle-\langle
n_{R,-\sigma}n_{R,\sigma} n_{L,\sigma} \rangle \nonumber \\
P_{2}&=&N_{R,\sigma}-\langle n_{R,\sigma} n_{L,\sigma}\rangle
-\langle n_{R,-\sigma} n_{R,\sigma}\rangle \nonumber \\ &+&\langle
n_{R,-\sigma} n_{R,\sigma} n_{L,\sigma}\rangle \nonumber \\
P_{3}&=&N_{R,-\sigma}-\langle n_{R,-\sigma} n_{L,\sigma}\rangle
-\langle n_{R,-\sigma}n_{R,\sigma}\rangle \nonumber \\ &+&\langle
n_{R,-\sigma}n_{R,\sigma} n_{L,\sigma} \rangle \nonumber \\
P_{4}&=&\langle n_{R,-\sigma}n_{R,\sigma}\rangle-\langle
n_{R,-\sigma}n_{R,\sigma} n_{L,\sigma}\rangle \nonumber\\
P_{5}&=&N_{L,\sigma}- \langle n_{R,\sigma}n_{L,\sigma}\rangle
-\langle n_{R,-\sigma} n_{L,\sigma}\rangle \nonumber \\ &+&\langle
n_{R,-\sigma}n_{R,\sigma} n_{L,\sigma}\rangle \nonumber \\
P_{6}&=&\langle n_{R,\sigma} n_{L,\sigma}\rangle -\langle
n_{R,-\sigma}n_{R,\sigma} n_{L,\sigma}\rangle \nonumber \\
P_{7}&=&\langle n_{R,-\sigma}n_{L,\sigma}\rangle -\langle
n_{R,-\sigma}n_{R,\sigma} n_{L,\sigma}\rangle \nonumber \\
P_{8}&=&\langle n_{R,-\sigma}n_{R,\sigma}n_{L,\sigma} \rangle
\nonumber,
\end{eqnarray}
\end{small}
where $N_{\ell,\sigma}$ represents the single-particle occupation
number of spin $\sigma$ at site $\ell$. The two-particle
correlation functions are denoted by $\langle n_{\ell,-\sigma}
n_{\ell,\sigma} \rangle$ (intra-dot) and $\langle n_{\ell,\sigma}
n_{j,\sigma} \rangle$ or $\langle n_{\ell,-\sigma} n_{j,\sigma}
\rangle$ (inter-dot), while the three-particle correlation
function is $\langle n_{\ell,-\sigma} n_{\ell,\sigma} n_{j,\sigma}
\rangle$. These correlation functions also have analytic
expressions derived in Ref.~[\onlinecite{Kuo3}]. All simulations
were performed using custom-developed Fortran code, utilizing IMSL
libraries for self-consistent Green's function iterations. Full
code is available upon request for reproducibility.



\setcounter{section}{0}
\setcounter{equation}{0} 

\mbox{}\\





\begin{thebibliography}{100}


\bibitem[1]{DiVincenzo} D. P. DiVincenzo, Quantum computations,
Science, 1995, \textbf{270}, 255.

\bibitem[2]{LloydS} S. Lloyd, Universal Quantum Simulators,
Science, 1996, \textbf{273}, 1073.

\bibitem[3]{ChuangIL} I. L. Chuang, R. Laflamme, P. Shor, and
W. H. Zurek, Quantum Computers, Factoring and Decoherence,
Science, 1995, \textbf{270}, 1633.

\bibitem[4]{SHOR} P. Shor, Scheme for reducing decoherence in quantum computer memory, Phys. Rev. A,  1995, \textbf{52}, R 2493.

\bibitem[5]{Loss} D. Loss and D. P. DiVincenzo, Quantum computation with quantum
dots, Phys. Rev. A, 1998, \textbf{57}, 120.

\bibitem[6]{Bennett} C. H. Bennett and  D. P. DiVincenzo, Quantum information and
computation, Nature, 2000, \textbf{404}, 247.

\bibitem[7]{Nakamura}  Y. Nakamura, Y. A. Pashkin, and
J. S. Tsai, Coherent control of macroscopic quantum states in a
single-Cooper-pair box, Nature, 1999, \textbf{398}, 786.

\bibitem[8]{Makhlin} Y. Makhlin, G. Schon, and A. Shnirman, Quantum-state engineering with Josephson-junction
devices, Rev. Mod. Phys., 2001, \textbf{73}, 357.

\bibitem[9]{Ono} K. Ono, D. G. Austing, Y. Tokura and S. Tarucha,
Current Rectification by Pauli Exclusion in a Weakly Coupled
Double Quantum Dot System, Science, 2002, \textbf{297}, 1313.

\bibitem[10]{vanderWiel} W. G. van der Wiel, S. De
Franceschi, J. M. Elzerman, T. Fujisawa, S. Tarucha, and L. P.
Kouwenhoven, Electron transport through double quantum dots, Rev.
Mod. Phys., 2003, \textbf{75}, 1.


\bibitem[11]{DeMille} D. DeMille, Quantum computation with trapped polar
molecules, Phys. Rev. Lett., 2002, \textbf{88}, 067901.

\bibitem[12]{Kielpinski} D. Kielpinski, C. Monroe, and D. J. Wineland, Architecture for a large-scale ion-trap
quantum computer, Nature, 2002, \textbf{417}, 709.

\bibitem[13]{DiVincenzoDP} D. P. DiVincenzo, Double quantum dot as a quantum
bit, Science, 2005, \textbf{309}, 2173.

\bibitem[14]{Petta} J. R. Petta, A. C. Johnson, J. M. Taylor, E.
A. Laird, A. Yacoby, M. D. Lukin, C. M. Marcus, M. P. Hanson and
A. C. Gossard, Coherent manipulation of coupled electron spins in
semiconductor quantum dots, Sceince,  2005, \textbf{309}, 2180.

\bibitem[15]{JphnsonAC} A. C. Johnson, J. R. Petta, C. M. Marcus,
M. P. Hanson and A. C. Gossard, Singlet-triplet spin blockade and
charge sensing in a few-electron double quantum dot, Phys. Rev. B,
2005, \textbf{72}, 165308.

\bibitem[16]{Koppens} F. H. L. Koppens, C. Buizert,
K. J. Tielrooij, I. T. Vink, K. C. Nowack, T. Meunier, L. P.
Kouwenhoven, and L. M. K. Vandersypen, Driven coherent
oscillations of a single electron spin in a quantum dot, Nature,
2006, \textbf{442}, 766.


\bibitem[17]{TaylorJM} J. M. Taylor, J. R. Petta,
A. C. Johnson, A. Yacoby, C. M. Marcus, and M. D. Lukin,
Relaxation, dephasing, and quantum control of electron spins in
double quantum dots, Phys. Rev. B., 2007, \textbf{76}, 035315.


\bibitem[18]{KoderaT} T. Kodera, K. Ono, Y.
Kitamura, Y. Tokura, Y. Arakawa, S. Tarucha, Quantitative
Estimation of Exchange Interaction Energy Using Two-Electron
Vertical Double Quantum Dots, Phys. Rev. Lett., 2009,
\textbf{102}, 146802.

\bibitem[19]{Maune} B. M. Maune, M. G. Borselli,
B. Huang, T. D. Ladd, P. W. Deelman, K. S. Holabird, A. A.
Kiselev,  I. Alvarado-Rodriguez, R. S. Ross, A. E. Schmitz, M.
Sokolich, C. A. Watson, M. F. Gyure, and A. T. Hunter, Coherent
singlet-triplet oscillations in a silicon-based double quantum
dot, Nature, 2012, \textbf{481}, 344.

\bibitem[20]{HaoXJ} X. J. Hao, R. Ruskov, M. Xiao, C. Tahan, and H. W. Jiang, Electron spin resonance and spin-valley physics in a silicon
double quantum dot, Nat. Commun., 2014, \textbf{5}, 3860.

\bibitem[21]{FrancescoR} F. Rossella, A. Bertoni, D. Ercolani, M.
Rontani, L. Sorba, F. Beltram and S. Roddaro, Nanoscale spin
rectifiers controlled by the Stark effect, Nat. Nanotechnol.,
2014, \textbf{9}, 997.

\bibitem[22]{WeberB} B. Weber, Y. H. M. Tan, S. Mahapatra, T. F. Watson,
H. Ryu, R. Rahman, L. C. L. Hollenberg,  G. Klimeck, and M. Y.
Simmons, Spin blockade and exchange in Coulomb-confined silicon
double quantum dots, Nat. Nanotechnol., 2014, \textbf{9}, 430.

\bibitem[23]{FujitaT} T. Fujita, P. Stano, G. Allison, K. Morimoto,
Y. Sato,  M. Larsson, J. H. Park, A. Ludwig, A. D. Wieck, and A.
Oiwa, Signatures of Hyperfine, Spin-Orbit, and Decoherence Effects
in a Pauli Spin Blockade, Phys. Rev. Lett., 2016, \textbf{117},
206802.

\bibitem[24]{ZhengH} H. Zheng, J. Y. Zhang, and R. Berndt, A minimal double quantum dot,
Sci. Rep., 2017, \textbf{7}, 10764.

\bibitem[25]{Zajac} D. M. Zajac, A. J. Sigillito, M. Russ, F.
Borjans, J. M. Taylor, G. Burkard and J. R. Petta, Resonantly
driven CNOT gate for electron spin, Science, 2018, \textbf{359},
439.

\bibitem[26]{TongC} C. Tong, A. Kurzmann, R. Garries, W. W. Huang,
S . Jele, M. Eich, L. Ginzgurg, C. Mittag and K. Watanable, Pauli
blockade of tunable two electron spin valley states in graphene
quantum dots, Phys. Rev. Lett.,  2022, \textbf{128}, 067702.


\bibitem[27]{MaRL} R. L. Ma, S. K. Zhu, Z. Z. Kong, T. P. Sun,
M. Ni, Y. C. Zhou, Y Zhou, G. Luo, G. Cao, G. L. Wang, H. O. Li,
and G. P. Guo, Singlet-triplet-state readout in silicon
metal-oxide-semiconductor double quantum dots, Phys. Rev. Appl.,
2024, \textbf{21}, 034022.


\bibitem[28]{Novoselovks} K. S. Novoselov, A. K. Geim, S. V. Morozov, D. Jiang, Y. Zhang,
S. V. Dubonos, I. V. Grigorieva, and A. A. Firsov, Electric Field
Effect in Atomically Thin Carbon Films, Science, 2004,
\textbf{306}, 666.

\bibitem[29]{Trauzettel} B. Trauzettel, D. V. Bulaev, D. Loss, and G. Burkard, Spin qubits in graphene quantum
dots, Nat. Phys., 2007, \textbf{3}, 192.

\bibitem[30]{AllenMT} M. T. Allen, J. Martin, and A. Yacoby, Gate-defined quantum confinement in suspended
bilayer graphene, Nat. commun., 2012, \textbf{3}, 934.

\bibitem[31]{Sarma} S. Das Sarma, S. Adam, E. H. Hwang, and E. Rossi, Electronic transport in two-dimensional graphene, Rev. Mod.
Phys., 2011, \textbf{83}, 407.

\bibitem[32]{Kotov} V. N. Kotov, B. Uchoa, V. M. Pereira, F. Guinea, and A. H.
Castro Neto, Electron-Electron Interactions in Graphene: Current
Status and Perspectives, Rev. Mod. Phys., 2012, \textbf{84}, 1067.

\bibitem[33]{VolkC} C. Volk, C.
Neumann, S. Kazarski, S. Fringes, S. Engels,  F. Haupt, A.
Mueller, and C. Stampfer, Probing relaxation times in graphene
quantum dots, Nat. commun., 2013, \textbf{4}, 1753.

\bibitem[34]{Brotons} M. Brotons-Gisbert,
B. Artur, S. Kumar, R. Picard, R. Proux, M. Gray, K. S. Burch, K.
Watanabe, T. Taniguchi, and B. D. Gerardot, Coulomb blockade in an
atomically thin quantum dot coupled to a tunable Fermi reservoir,
Nat. nanotechonol., 2019, \textbf{14}, 442.

\bibitem[35]{Cai} J. Cai, P. Ruffieux, R. Jaafar, M. Bieri, T. Braun, S. Blankenburg, M. Muoth,
A. P. Seitsonen, M. Saleh, X. Feng, K. Mullen, and Roman Fasel,
Atomically precise bottom-up fabrication of graphene nanoribbons,
Nature, 2010, \textbf{466}, 470.


\bibitem[36]{ChenYC} Y. C. Chen, T. Cao, C. Chen, Z. Pedramraz, D.
Haberer, D. G. de Oteyza, F. R. Fischer, S. G. Louie and M. F.
Crommie, Molecular bandgap engineering of bottom-up synthesized
graphene nanoribbon heterojunctions, Nat. Nanotechnol., 2015,
\textbf{10}, 156.

\bibitem[37]{WangS} S. Y. Wang, L. Talirz, Carlo A. Pignedoli, X. L. Feng, K. Mullen, R.
Fasel and P. Ruffieux, Giant edge state splitting at atomically
precise graphene zigzag edges, Nat. Commun., 2015, \textbf{7},
11507.

\bibitem[38]{LlinasJP} J. P. Llinas,  A. Fairbrother,  Barin G. Borin, W. Shi, K. Lee, S. Wu,
B. Y. Choi, R. Braganza, J. Lear and N. Kau, Short-channel
field-effect transistors with 9-atom and 13-atom wide graphene
nanoribbons, Nat. Commun., 2017, \textbf{8}, 633.

\bibitem[39]{Nestor} N. Merino-Diez, A. Garcia-Lekue, E. Carbonell-Sanroma,`
J. C. Li, M. Corso, L. Colazzo, F. Sedona, D. Sanchez-Portal, J.
I. Pascual, and Dimas G. de Oteyza, Width-Dependent Band Gap in
Armchair Graphene Nanoribbons Reveals Fermi Level Pinning on
Au(111), ACS nano, 2017, \textbf{11}, 11661.

\bibitem[40]{Groning} O. Groning, S. Wang, X. Yao, C.
A. Pignedoli, G. B. Barin, C. Daniels, A. Cupo, V. Meunier, X.
Feng, A. Narita, et al., Engineering of robust topological quantum
phases in graphene nanoribbons, Nature, 2018, \textbf{560}, 209.

\bibitem[41]{Rizzo} D. J. Rizzo, G. Veber, T. Cao, C. Bronner, T. Chen,
F. Zhao, H. Rodriguez, S. G. Louie, M. F. Crommie, and F. R.
Fischer, Topological band engineering of graphene nanoribbons,
Nature, 2018, \textbf{560}, 204.

\bibitem[42]{Yan}L. H. Yan and P. Liljeroth, Engineered electronic states in
atomically precise artificial lattices and graphene nanoribbons,
Advances in Physics: X, 2019, \textbf{4}, 1651672.

\bibitem[43]{DRizzo} D. J. Rizzo, G. Veber, J. W. Jiang, R. McCurdy, T. Cao
C. Bronner, T. Chen, Steven G. Louie1, F. R. Fischer, and M. F.
Crommie, Inducing metallicity in graphene nanoribbons via
zero-mode superlattices, Science,  2020, \textbf{369}, 1597.

\bibitem[44]{SunQ} Q. Sun, Y. Yan, X. L. Yao, K. Mullen, A. Narita, R. Fasel, and P. Ruffieux,
Evolution of the topological energy band in graphene Nanoribbons,
J. Phys. Chem. Lett., 2021, \textbf{12}, 8679.

\bibitem[45]{DJRizzo} D. J. Rizzo, J. W. Jiang, D. Joshi, G. Veber, C. Bronner, R. A. Durr,
P. H. Jacobse, T. Cao, A. Kalayjian, H. Rodriguez, P. Butler, T.
Chen, Steven G. Louie, F. R. Fischer, and M. F. Crommie,
Rationally designed topological quantum dots in bottom-up graphene
nanoribbons, ACS Nano, 2021, \textbf{15}, 20633.

\bibitem[46]{SongST} S. T Song, Y. Teng, W. C. Tang, Z. Xu, Y. Y. He; J. W. Ruan, T.
Kojima, W. P. Hu, F. J. Giessibl, H. Sakaguchi,
 S. G. Louie, and J. Lu, Janus graphene nanoribbons with localized
states on a single zigzag edge, Nature, 2025, \textbf{637}, 580.

\bibitem[47]{Albrecht} S. M. Albrecht, A. P
Higginbotham, M. Madsen, F. Kuemmeth, T. S. Jespersen, J. Nygard,
P. Krogstrup, and C. M. Marcus, Exponential protection of zero
modes in Majorana islands, Nature, 2016, \textbf{531}, 201.

\bibitem[48]{BorsoiF}F. Borsoi, N. W. Hendrickx, V. John, M. Meyer, S. Motz, F. van
Riggelen, A. Sammak, Sander L. de Snoo, G. Scappucci and M.
Veldhorst, Shared control of a 16 semiconductor quantum dot
crossbar array, Nat. Nanotechnol., 2024, \textbf{19}, 21.

\bibitem[49]{WangHM} H. M. Wang, H. S. Wang, C. X. Ma, L. X. Chen, C. X. Jiang, C.
Chen, X. M. Xie, A. P. Li and X. R. Wang, Graphene nanoribbons for
quantum electronics, Nat. Rev. Phys., 2021, \textbf{3}, 791.

\bibitem[50]{LiangGC} G. C. Liang, N. Neophytou, M. S. Lundstrom
and D. E. Niknonov, Contact effects in graphene nanoribbon, Nano
Lett., 2008, \textbf{8}, 1819.

\bibitem[51]{ThiagoB} T. B. Martins, A. J. R. da Silva, R. H. Miwa
and A. Fazzio, $\sigma-$ and $\pi$-Defects at graphene nanoribbon
edges: building spin filters, Nano Lett., 2008, \textbf{8}, 2293.

\bibitem[52]{ChenRS} R. S. Chen, G. L. Ding, Y. Zhou, and S. T.  Han, Fermi-level depinning
of 2D transition metal dichalcogenide transistors. J. Mater. Chem.
C, 2021, \textbf{9}, 11407.


\bibitem[53]{Mangnus} M. J. J. Mangnus, F. R. Fischer, M. F. Crommie, I. Swart and P. H.
Jacobse, Charge transport in topological graphene nanoribbons and
nanoribbon heterostructures, Phys. Rev. B, 2022, \textbf{105},
115424.

\bibitem[54]{Kuo1} David. M. T. Kuo and Y. C. Chang, Contact Effects on Thermoelectric Properties of Textured Graphene
Nanoribbons, Nanomaterials, 2022, \textbf{12}, 3357.


\bibitem[55]{Zdetsis} A. D. Zdetsis, Peculiar electronic properties of wider armchair
graphene nanoribbons: Multiple topological end-states and "phase
transitions", Carbon, 2023, \textbf{210}, 118042.

\bibitem[56]{Kuo2} D. M. T. Kuo, Charge transport through the multiple end zigzag edge states of armchair graphene nanoribbons and
heterojunctions, RSC Adv., 2024, \textbf{14}, 20113.

\bibitem[57]{YouSF} S. F. You, C. J. Yu, Y. X. Gao, X. C. Li, G.
Y. Peng, K. F. Niu, J. H. Xi, C. J. Xu, S. X. Du, X. X. Li, J. L.
Yang and L. F. Chi, Quantifying the conductivity of a single
polyene chain by lifting with an STM tip, Nat. commun., 2024,
\textbf{15}, 6475.

\bibitem[58]{HuangAH} A. H. Huang, S. S. Ke, J. H. Guan, J. Li, and W. K, Lou,
Floquet-Engineering Topological Phase Transition in Graphene
Nanoribbons by Light, Chin. Phys. Lett., 2024, \textbf{41},
097302.

\bibitem[59]{GuYan} Y. W. Gu, Z. Qiu and K. Mullen, Nanographenes
and Graphene Nanoribbons as Multitalents of Present and Future
Materials Science, J. Am. Chem. Soc., 2022, \textbf{144} 11499.

\bibitem[60]{Golor} M. Golor, C. Koop, T. C. Lang, S. Wessel and
M. J. Schmidt, Magnetic correlations in short and narrow graphene
armchair nanoribbons, Phys. Rev. Lett., 2013, \textbf{111},
085504.

\bibitem[61]{ChenCC} C. C. Chen and Y. C. Chang, Theoretical studies of graphene nanoribbon quantum dot qubits,
Phys. Rev. B, 2015, \textbf{92}, 245406.

\bibitem[62]{BanY} Y. Ban, K. Kato, S. Iizuka, H. Oka, S. Murakami,
K. Ishibashi, S. Moriyama, T. Mori, K. Ono, Pauli spin blockade at
room temperature in double-quantum-dot tunneling transport through
individual deep dopants in silicon, Commun. Phys. 2025,
\textbf{8}, 293.

\bibitem[63]{Nakada} K. Nakada, M. Fujita, G. Dresselhaus and M. S. Dresselhaus,
Edge state in graphene ribbons: Nanometer size effect and edge
shape dependence, Phys. Rev. B, 1996, \textbf{54},  17954.

\bibitem[64]{Wakabayashi} K. Wakabayashi, M. Fujita, H. Ajiki, and M. Sigrist,
Electronic and magnetic properties of nanographite ribbons, Phys.
Rev. B, 1999, \textbf{59}, 8271.

\bibitem[65]{Wakabayashi2} K. Wakabayashi, K Sasaki,
T. Nakanishi and T. Enoki, Electronic states of graphene
nanoribbons and analytical solutions, Sci. Technol. Adv. Mater.,
2010, \textbf{11}, 054504.

\bibitem[66]{SunQF} Q. F. Sun and X. C. Xie, CT-Invariant Quantum Spin Hall Effect in
Ferromagnetic Graphene, Phys. Rev. Lett., 2010, \textbf{104},
066805.

\bibitem[67]{SSH}  W. P. Su,  J. R. Schrieffer and  A. J. Heeger, Soliton excitations
in polyacetylene, Phys. Rev. B, 1980, \textbf{22}, 2099.

\bibitem[68]{LieuS} S. Lieu, Topological phases in the non-Hermitian Su-Schrieffer-Heeger
model, Phys. Rev. B, 2018, \textbf{97}, 045106.

\bibitem[69]{ObanaD}  D. Obana, F. Liu and  K. Wakabayashi, Topological edge states in the
Su-Schrieffer-Heeger model, Phys. Rev. B, 2019, \textbf{100}
075437.

\bibitem[70]{SanchoMP} M. P. Lopez-Sancho and M. Carmen Munoz, Topologically protected
edge and confined states in finite armchair graphene nanoribbons
and their junctions, Phys. Rev. B, 2021, \textbf{104}, 245402.

\bibitem[71]{Kuo4} David M. T. Kuo, Topological states in finite graphene nanoribbons tuned by
electric fields, J. Phys.: Condens. Matter, 2025, \textbf{37},
085304.

\bibitem[72]{MatsudaY} Y. Matsuda,  W. Q. Deng,  W. A. Goddard III, Contact Resistance
for "End-Contacted" Metal-Graphene and Metal-Nanotube Interfaces
from Quantum Mechanics, J. Phys. Chem. C, 2010,  \textbf{114},
17845.

\bibitem[73]{OtsukaT} T. Otsuka, T. Nakajima, M. R. Delbecq, S. Amaha,
J. Yoneda, K. Takeda, G. Allison, T. Ito, R. Sugawara, A. Noiri,
A. Ludwig, A. D. Wieck and Seigo Tarucha, Single-electron Spin
Resonance in a Quadruple Quantum Dot, Sci. Rep., 2016, \textbf{6},
31820.

\bibitem[74]{GaoQ} Q. Gao and  J. Guo, Role of chemical termination in edge contact to
graphene, APL Mater., 2014, \textbf{2} 056105.

\bibitem[75]{Falicov} L. M. Falicov and R. A. Harris, Two-electron homoploar molecule: a test for spin denisty
waves and charge density waves, J. Chem. Phys., 1969, \textbf{51},
3153.

\bibitem[76]{PizzocheroM} M. Pizzochero, N. V. Tepliakov, J.
Lischner, A. A. Mostofi, and E. Kaxiras, One-Dimensional Magnetic
Conduction Channels across Zigzag Graphene Nanoribbon/ Hexagonal
Boron Nitride Heterjunctions, Nano Lett., 2024, \textbf{24}, 6521.

\bibitem[77]{TepliakovNV} N. V. Tepliakov, R. Ma, J. Lischner, E.
Kaxiras, A. A. Mostofi, and M. Pizzochero, Dirac Half-Semimetally
and Antiferromagnectism in Graphene Nanoribbon/Hexagonal Boron
Nitride Heterojunctions, Nano Lett., 2023, \textbf{23}, 6698.

\bibitem[78]{FranssonJ} J. Fransson and M. Rasander, Pauli spin blockade in weakly coupled double
quantum dots, Phys. Rev. B, 2006, \textbf{73}, 205333.

\bibitem[79]{Muralidharan} B. Muralidharan and S. Datta, Generic model for current collapse in
spin blockaded transport, Phys. Rev. B, 2007, \textbf{76}, 035432.

\bibitem[80]{Inarrea} J. Inarrea, G. Platero and A. H. MacDonald, Electronic transport through a
double quantum dot in the spin-blockade regime:Theorical models,
Phys. Rev. B, 2007, \textbf{76}, 085329.

\bibitem[81]{DanonJ} J. Danon and Yu. V. Nazarov, Pauli spin blockade in the presence of strong spin-orbit
coupling, Phys. Rev. B, 2009, \textbf{80}, 041301 (R).

\bibitem[82]{PalyiA} A. Palyi and G. Burkard, Hyperfine-induced valley mixing and the spin-valley blockade in carbon-based quantum
dots, Phys. Rev. B, 2009, \textbf{80}, 201404 (R).

\bibitem[83]{Kuo5} David. M. T. Kuo, S. Y. Shiau and Y. C. Chang, Theory of spin blockade, charge
ratchet effect, and thermoelectrical behavior in serially coupled
quantum dot system, Phys. Rev. B, 2011, \textbf{84}, 245303.

\bibitem[84]{DasSarma} D. Das Sarma, X. Wang, and S. Yang, Hubbard model description of
silicon spin qubits: Charge stability diagram and tunnel coupling
in Si double quantum dots, Phys. Rev. B, 2011, \textbf{83},
235314.

\bibitem[85]{DasSarma1} X. Wang, S. Yang, and S. Das Sarma, Quantum theory of the
charge-stability diagram of semiconductor double-quantum-dot
systems, Phys. Rev. B, 2011, \textbf{84}, 115301.

\bibitem[86]{HouWJ} W. J. Hou, Y. D. Wang, J. H. Wei, and Y. J.
Yan, Manipulation of Pauli spin blockade in double quantum dot
systems, J. Chem. Phys., 2017, \textbf{146}, 224304.

\bibitem[87]{Kondo} M. Kondo, S. Miyota, W. Izumida, S. Amaha, and T.
Hatano, Thermally assisted Pauli spin blockade in double quantum
dots, Phys. Rev. B, 2021, \textbf{103}, 155414.

\bibitem[88]{Kuo3} David. M. T. Kuo, Temperature-stable tunneling current in serial double quantum dots:
insights from nonequilibrium Green's functions and Pauli spin
blockade, Phys. Chem. Chem. Phys., 2025, \textbf{27}, 5238.

\bibitem[89]{LandiGT} G. T. Landi, D. Polettic and G. Schaller,
Nonequilibrium boundary-driven quantum systems: Models, methods,
and properties, Rev. Mod. Phys., 2024, \textbf{94}, 045006.

\bibitem[90]{JeongH} H. Jeong, A. M. Chang, and M. R.
Melloch, The Kondo effect in an artificial quantum dot molecule,
Science, 2001, \textbf{293}, 2221.

\bibitem[91]{AguadoR} R. Aguado, and D. C. Langreth,
Out-of-equilibrium Kondo effect in double quantum dots, Phys. Rev.
Lett., 2000, \textbf{85}, 1946.

\bibitem[92]{ChungCH} C. H. Chung and T. H. Lee, Tunable Fano-Kondo resonance in side-coupled double quantum dot
systems, Phys. Rev. B, 2010, \textbf{82}, 085325.

\bibitem[93]{WangJN} J. N. Wang, Y. C.
Xiong, W. H. Zhou, T. Peng and Z. Y. Wang, Secondary proximity
effect in a side-coupled double quantum dot structure, Phys. Rev.
B, 2024, \textbf{109}, 064518.

\bibitem[94]{BorselliMG} M. G. Borselli, K. Eng, E. T. Croke, B. M. Maune, B. Huang, R. S.
Ross, A. A. Kiselev, P. W. Deelman, I. Alvarado-Rodriguez, A. E.
Schmitz, M. Sokolich, K. S. Holabird, T. M. Hazard, M. F. Gyure
and A. T. Hunter, Pauli spin blockade in undoped Si/SiGe
two-electron double quantum dots, Appl. Phys. Lett., 2011,
\textbf{99}, 063109.

\bibitem[95]{ZhangJ} J. Zhang, O. Braun, G. B. Barin, S. Sangtarash,
J. Overbeck, R. Darawish, M. Stiefel, R. Furrer, A. Olziersky, K.
Mullen, I. Shorubalko, A. H. S. Daaoub, P. Ruffieux, R. Fasel, H.
Sadeghi, M. L. Perrin, and M. Calame, Tunable Quantum Dots from
Atomically Precise Graphene Nanoribbons Using a Multi-Gate
Architecture, Adv. Electron. Mater., 2023, \textbf{9}, 2201204.

\bibitem[96]{MariusE} M. Eich, R. Pisoni, A. Pally, H. Overweg, A. Kurzmann, Y. J. Lee,
P. Rickhaus, K. Watanabe, T. Taniguchi, K. Ensslin, and T. Ihn,
Coupled Quantum Dots in Bilayer Graphene, Nano Lett., 2018,
\textbf{18}, 5042.

\bibitem[97]{Abbassi} M. E. Abbassi, M. L. Perrin, G. B. Barin, S. Sangtarash, J. Overbeck,
O. Braun, C. J. Lambert, Q. Sun, T. Prechtl, A. Narita, K. Mullen,
P. Ruffieux, H. Sadeghi, R. Fasel, and M. Calame, Controlled
Quantum Dot Formation in Atomically Engineered Graphene Nanoribbon
Field-Effect Transistors, ACS Nano,  2020, \textbf{14}, 5754.

\bibitem[98]{Huangwh} W. H. Huang, O. Braun, David I. Indolese, G.
B. Barin, G. Gandus, M. Stiefel, A. Olziersky, K. Mullen, M.
Luisier, D. Passerone, P. Ruffieux, C. Schonenberger, K. Watanabe,
T. Taniguchi, R. Fasel, J. Zhang, Michel Calame, and M. L. Perrin,
Edge Contacts to Atomically Precise Graphene Nanoribbons, ACS
Nano, 2023, \textbf{17}, 18706.

\bibitem[99]{ZhangJain} J. Zhang et al, Contacting individual
graphene nanoribbons using carbon nanotube electrodes, Nat.
electron., 2023, \textbf{6}, 572.


\bibitem[100]{TaoNJ} N. J. Tao, Electron transport in molecular
junctions, Nature Nanotechnology, 2006, \textbf{1}, 173.

\bibitem[101]{AndrewsDQ} D. Q. Andrews, G. C. Solomon, R. P. Van Duyne, and M. A. Ratner,
Single Molecule Electronics: Increasing Dynamic Range and
Switching Speed Using Cross-Conjugated Species, J. Am. Chem. Soc.,
2008, \textbf{130}, 17309.

\bibitem[102]{Kuo7} David M. T. Kuo, Thermal rectification through the topological states of
asymmetrical length armchair graphene nanoribbons heterostructures
with vacancies, Nanotechnology, 2023, \textbf{34}, 505401.

\bibitem[103]{LuoYi} Y. Luo et al, One-Dimensional Quantum Dot
Array Integrated with Charge Sensors in an InAs Nanowire, Nano
Lett., 2024,\textbf{ 24}, 14012.




















\end{thebibliography}
\end{document}